# A rhombic spin-1 interpretation of a room-temperature GaN emitter

Daniel Hashemi
*Department of Physics, Optical Engineering and Nanoengineering, Rose-Hulman Institute of Technology, Terre Haute, Indiana 47803, USA*
hashemi@rose-hulman.edu

**Abstract.** Single defects in GaN show optically detected magnetic resonance at room temperature. In the more abundant class, a minimal $S=3/2$ model leaves a $g\approx4$ line unexplained. For one defect, this line lies near the sum of the two outer lines, consistent with transitions among three shared levels. A rhombic $S=1$ model reproduces the three line positions with rms errors of 8.1–9.8 MHz. The tested second-order $S=3/2$ fits containing all three transitions and the published minimal model do not reproduce these positions. A ladder fitted to the strong lines remains viable if the $g\approx4$ line is a separate transition near the sum of the outer two. An unresolved frequency-window shift prevents the existing data from establishing the $S=1$ assignment uniquely. We propose a zero-field sweep that directly tests the isolated $S=1$ assignment.

## I. Introduction

Single defects in GaN, a semiconductor used in power and high-frequency electronics, show optically detected magnetic resonance (ODMR) at room temperature [1,2], as do defects in other wide-gap hosts [3,4]. Luo *et al.* identified two classes of emitters. The more abundant class, group II, shows the largest contrast, reaching 30 % [1]. Its spin and microscopic origin remain unresolved [5–8].

Luo *et al.* described the three strong lines of group II, $u$ (upper), $c$ (central) and $l$ (lower), using an $S=3/2$ spin Hamiltonian with $D=368$ MHz and $E=0$. They presented this as a minimal model and concluded that $S\geq3/2$ [1]. They also reported a fourth line, $q$, with a field dispersion corresponding to $g=4$, which the minimal model does not explain. Here we reanalyze their measurements using the raw data and analysis code they made public [9].

A single emitter can exhibit both $S=1$ and $S=1/2$-like resonances. Robertson *et al.* showed how charge transfer between two defects can produce this behavior and identified GaN as a possible example in their preprint [8,10,11]. Whitefield *et al.* observed such coexistence in hBN [12]. In their preprint, Whitefield *et al.* assigned a $g=4$ line to the $S=1$ double-quantum transition [13]; their published Supplement instead favors the $T_+\leftrightarrow T_-$ transition of a weakly coupled pair [12,14]. These studies establish precedents for both the coexistence of resonances and a $g\approx4$ line in a three-level system.

Here we focus on what the line positions reveal. In defect 2 of the public data [9], $q$ lies near the sum of the two outer lines: $q-(u+l)$ falls between $-43$ and $+16$ MHz in a 95% mixture interval, assuming a constant shift of $l$ and equal weights for assigning that shift to either frequency window. By comparison, $q$ lies 102 MHz from twice the central line. For this defect, the tested second-order $S=3/2$ fits containing $q$ fail to reproduce the positions, as does the minimal model of Luo *et al.* A second-order $S=3/2$ description remains possible: a ladder for $u$, $c$ and $l$ accompanied by an unrelated $q$ line that happens to fall at $u+l$. A strongly rhombic $S=1$ manifold accounts for $u$, $l$ and $q$ without this coincidence, while treating the broad central line $c$ as a separate spin, charge configuration or center. Table I summarizes the competing models and proposed tests. A zero-field sweep tests the isolated $S=1$ assignment, while two-tone ODMR probes a spin pair hidden under $c$.

## II. The combination rule

The sum of two transition frequencies can reveal whether three lines connect the same three levels. For any three levels $E_0 < E_1 < E_2$, the transition frequencies obey $f_{02} = f_{01} + f_{12}$, the Ritz combination principle. For an $S = 1$ spin above the level crossing, the upper and lower single-quantum lines $u$ and $l$ share the middle level. The double-quantum line $q$ connects the two outer levels, so its frequency is their sum,

$$q = u + l. \quad (1)$$

We use the spin Hamiltonian

$$H = D'\left(S_z^2 - 2/3\right) + E\left(S_x^2 - S_y^2\right) + g\,\mu_B\,B \cdot S \quad (2)$$

where $D'$ and $E$ are the axial and rhombic zero-field parameters and $\mu_B$ is the Bohr magneton. For $B \parallel z$, the line frequencies are $f_u = D' + s$, $f_l = |D' - s|$ and $f_q = 2s$, where $s = \sqrt{E^2 + (g\,\mu_B\,B)^2}$ is half the splitting between the two levels derived from $m_s = \pm 1$. Equation (1) holds above $B_c \approx 27$ mT (272 G on the field scale of the data), where $s > D'$. Below $B_c$, the same levels instead give $q = u - l$ [14].

This relation constrains the connections among three levels without fixing the total spin. In the $S = 3/2$ ladder proposed by Luo *et al.*, $l$, $c$ and $u$ connect adjacent levels [Fig. 1(a)]. In that ladder, $u + l$ does not correspond to a transition. For an axial $S = 3/2$ spin with the field along its axis, $u + l = 2c$ [14]. A transition at $u + l \neq 2c$ is therefore inconsistent with that axial ladder. However, three of the four $S = 3/2$ levels can still support $l$, $u$ and $q = u + l$. The combination rule thus points to a three-level manifold, for which $S = 1$ is the minimal isolated-spin realization. Larger manifolds must be tested through their other predictions.

## III. Data

The public data [9] include continuous-wave (cw) ODMR field sweeps for defects 1–4, recorded in 4, 2, 2 and 1 frequency windows, respectively. They also contain pulsed ODMR, microwave-on/off fluorescence transients, cw traces near 1 kG and angle scans for defects 1 and 2. The archive also contains photon correlations, spectra, fluorescence maps and magneto-photoluminescence, along with Rabi data for defect 6.

We analyze combinations of lines within each field column, which is the spectrum recorded at one field value. These combinations are independent of the field scale. We fit each column with a sum of lines, include the amplifier replicas at $c/2$ and $u/2$ [1], and exclude columns in which $l$ lies within 80 MHz of either replica [14].

## IV. Results

**TABLE I.** Tests of the competing interpretations for defect 2. Unless stated otherwise, the tests use line positions. Details are given in the text and Ref. [14].

| Hypothesis or alternative | Test | Assumptions or requirements | Outcome |
|---|---|---|---|
| $q$ is noise | Track statistic of $q$ | Null maps preserving strong lines, replicas and correlated noise | Detected in all 21 columns; no null map reaches one third of its track statistic |
| $q$ at $2c$ ($\Delta m_s = 2$ line of the spin of $c$; axial $S = 3/2$) | $q - 2c$ in every column | Window offset | $q - 2c \geq 58$ MHz; removing this difference requires an offset excluded at $4.4\,\sigma$ |
| Weak pair hidden in $c$ ($T_+ \leftrightarrow T_-$) | $q$ compared with twice the centroid of the pair's lines | Contrast weights equal to line strengths | Disfavored, not excluded; test (iv) checks it |

| Hypothesis or alternative | Test | Assumptions or requirements | Outcome |
|---|---|---|---|
| The best $S=3/2$ fitted to $u$, $c$ and $l$ | Fit $u$, $c$, $l$; predict $q$ | Free $D$, $E$, orientation, anisotropic $g$ | Differs from $q$ by at least $321$ MHz |
| $S=3/2[u,c,l]$ plus an independent line at $u+l$ | Calibrated test and parsimony | Constant window shift of $l$; assumption that the coincidence is unlikely | Inconclusive if the shift is allowed to vary with field; one error model mildly favors the ladder |
| $S=3/2[u,l,q]$, $c$ separate | Joint and out-of-sample fits | Second-order spin Hamiltonian | $\Delta\chi^2=7942$; rejected by the line positions within the tested models |
| $S=1+c$ | Joint fit; $q$ from $u$ and $l$ alone | Stated errors; no correction for the window shift | $\chi^2/dof=1.28$, a poor fit at the stated errors; $q$ predicted to $10\pm4$ MHz rms |
| Isolated $S=1$ | Zero-field sweep (proposed) | $u$ and $l$ tracked continuously from high field | Untested; lines approaching zero or splitting linearly would falsify the model |

### A. Recovering the weak line

The line $q$ reported by Luo *et al.* is visible in all $21$ columns from $331$ to $705$ G [Fig. 1(b)]. We compared it with residual-bootstrap null maps that omit $q$ while retaining the strong lines, amplifier replicas and correlated noise. None reaches one third of the observed track statistic; the family-wise p-value is 0.010–0.024 over all 45 search statistics [14]. The curvature is detected at $3.4$–$4.6\sigma$ when $c$ serves as the field reference or a shared field distortion is allowed. The significance reaches $6.1\sigma$ only when measured against the nominal field [14]. This curvature is consistent with $f_q=2s$, but a separate curved line would also fit.

### B. The line lies near $u+l$

The closure combines measurements from two frequency windows, so any shift between them enters the result directly. All four lines are detected in $8$ columns between $560$ and $705$ G. In these columns, $l$ is recorded in the lower window, whereas $u$ and $q$ are recorded in the upper window. Where $l$ appears in both windows, at $895$–$997$ G, its fitted positions differ by $31\pm8$ MHz. The $u/2$ replica in the same columns shows no such shift [14]. The available data do not establish which window carries the shift. Giving equal prior weight to the two constant-offset possibilities gives

$$q-(u+l)=-11.1\pm16.7\ MHz \quad (3)$$

with a central 95% mixture interval from $-43$ to $+16$ MHz, assuming a constant offset. If the shift belongs to the lower window, $q-(u+l)=-24.8\pm10.9$ MHz, which is $2.3\sigma$ from zero. If it belongs to the upper window, the result is $+2.8\pm7.9$ MHz [14]. The closure also drifts with field, as discussed in Sec. IV G. Equation (3) tests whether the transitions share three levels; it does not by itself determine the spin.

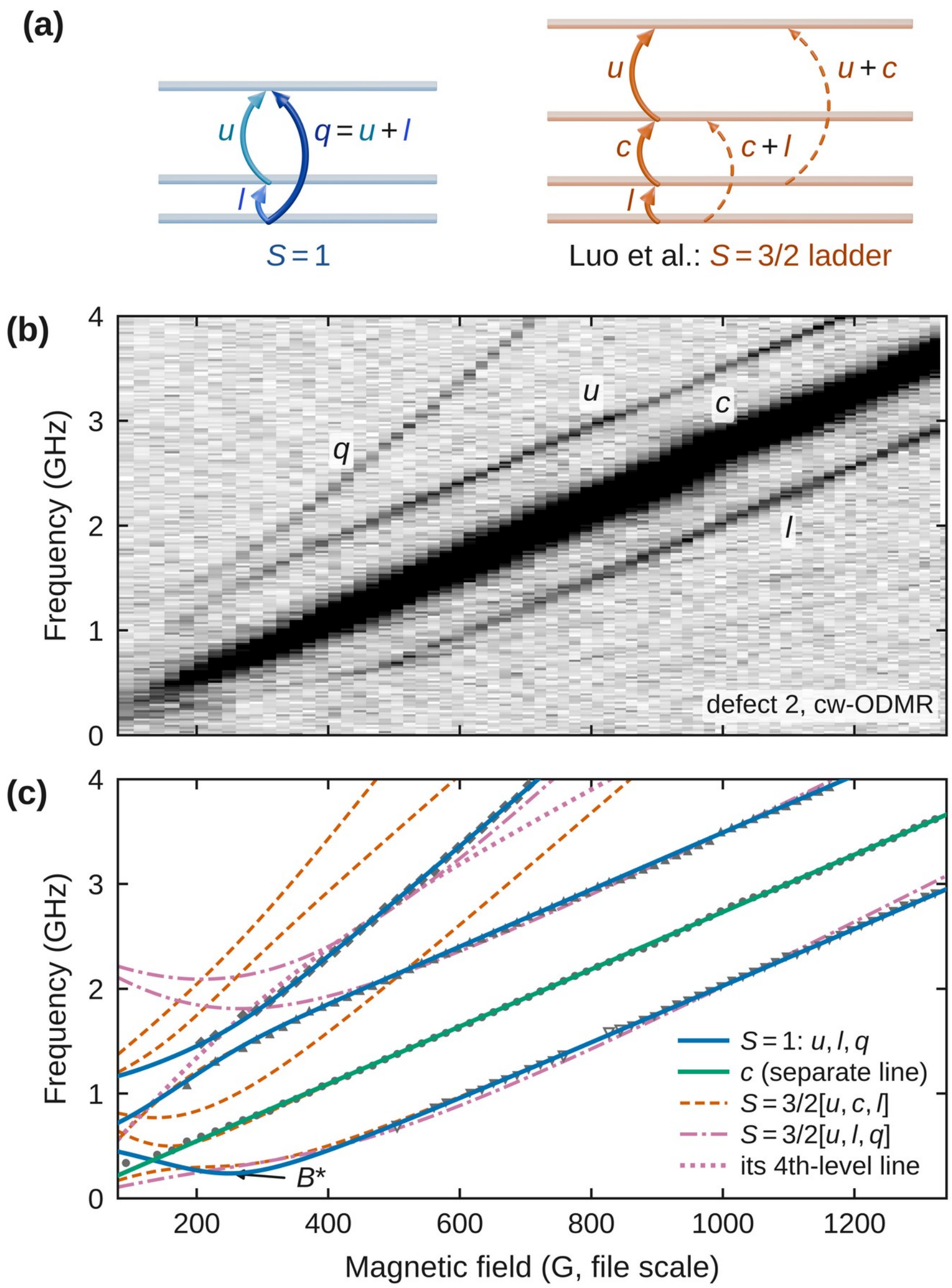

**FIG. 1.** (a) Level schemes at $633$ G, drawn to scale. For $S=1$, $q$ connects the outer levels. In the $S=3/2$ ladder of Luo *et al.*, $l$, $c$ and $u$ connect adjacent levels; dashed lines mark the $\Delta m_s=2$ sums. (b) cw-ODMR contrast of defect 2 from the public data [9] (CC BY-NC 4.0), shown with a saturated color scale to reveal $q$. (c) Fitted line positions in grey, with open symbols where $l$ blends with $c/2$. Curves show the $S=1+c$ model (solid), one best $S=3/2[u,c,l]$ fit (dashed), and the $S=3/2[u,l,q]$ fit with free $g$ (dash-dotted). Its fourth-level line is dotted and is not used as evidence. The predicted avoided crossing is marked by $B^*$.

A $\Delta m_s=2$ transition of the spin responsible for $c$, or the axial $S=3/2$ identity, would place $q$ at $2c$. Instead, $q-2c$ is at least $58$ MHz in every column. Removing this difference would require a window offset beyond $-10$ G, which is excluded at $4.4\sigma$ [14]. Another possibility is the $T_+ \leftrightarrow T_-$ transition of a spin pair whose $\Delta m_s=1$ lines are hidden within the broad $c$ line. For a weak pair, however, the transition lies within 5.0 MHz of twice the line-strength centroid of those lines, unless their contrast weights differ from their line strengths [14]. Two-tone ODMR [test (iv)] can check the pair hypothesis directly.

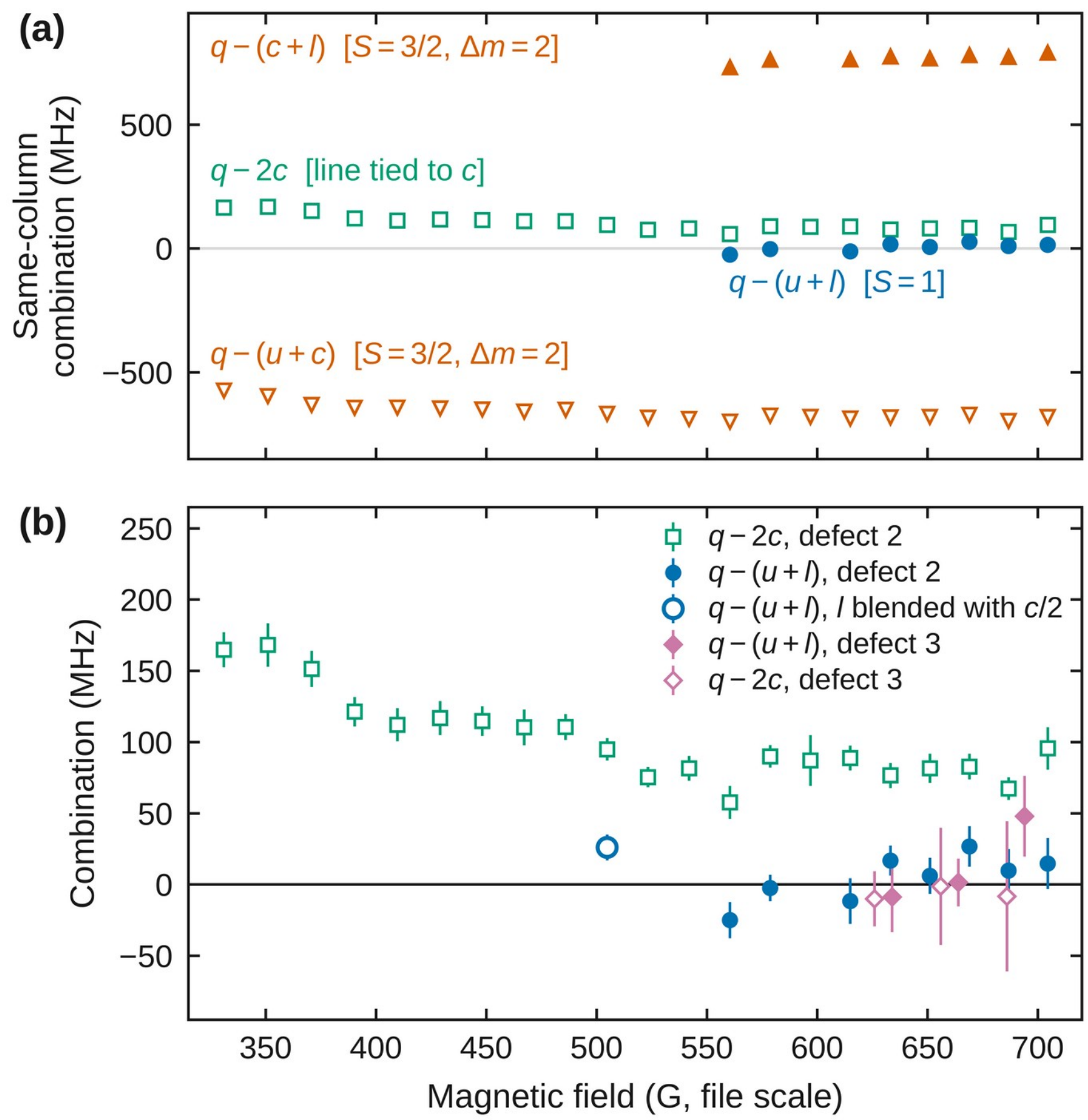

**FIG. 2.** The line $q$ lies near $u+l$ and is separated from $2c$, $u+c$ and $c+l$. (a) For defect 2, filled circles show $q-(u+l)$ without a window correction. They would shift downward by about 29 MHz if the lower window carries the shift of $l$. Open squares show $q-2c$; triangles show the differences from the $\Delta m_s=2$ sums, $q-(u+c)$ and $q-(c+l)$. (b) Enlarged view. The open circle marks the excluded 505 G column, where $l$ blends with $c/2$. Diamonds show defect 3 under relaxed detection criteria, displaced for clarity. Error bars are $\pm 1\sigma$.

### C. The $u$, $c$, $l$ ladder does not reproduce $q$

The best $S=3/2$ model fitted to $u$, $c$ and $l$ allows $D$, $E$, the orientation and the anisotropic $g$ tensor to vary. Its predictions differ from $q$ by at least 321 MHz in every column, even after including $\Delta m_s=2$ sums and multiphoton lines [Figs. 2(a) and 3(a)]. An $S=3/2$ model forced to contain all four lines requires an effective $g$ of 1.45 and still gives a worse fit [14].

The ladder remains compatible with an independent line at $u+l$. In a calibrated test of the line positions, with $q$ free in the ladder and both models refitted, this interpretation is rejected at $p\le 0.0075$ if the window shift of $l$ is a constant offset. Measurements do not establish that assumption, and the drift of the closure does not support it. Under the same assumption, the $S=1$ + $c$ model gives $p\ge 0.04$, which is marginal in one case. If the shift is allowed to vary with field, the test is inconclusive and one error model mildly favors the ladder. The positions of $u$ and $c$ alone disfavor it only weakly ($p=0.010$–$0.075$) [14]. Without assuming a constant offset, the preference for spin-1 rests on parsimony and on the assumption that an independent line is unlikely to follow $u+l$. Prior-dependent coincidence calculations are given in Ref. [14]; they do not resolve the window-shift ambiguity. The low-field positions of $c$ disfavor the ladder only if the field direction remains fixed, and that preference depends mainly on four uncertain points [14].

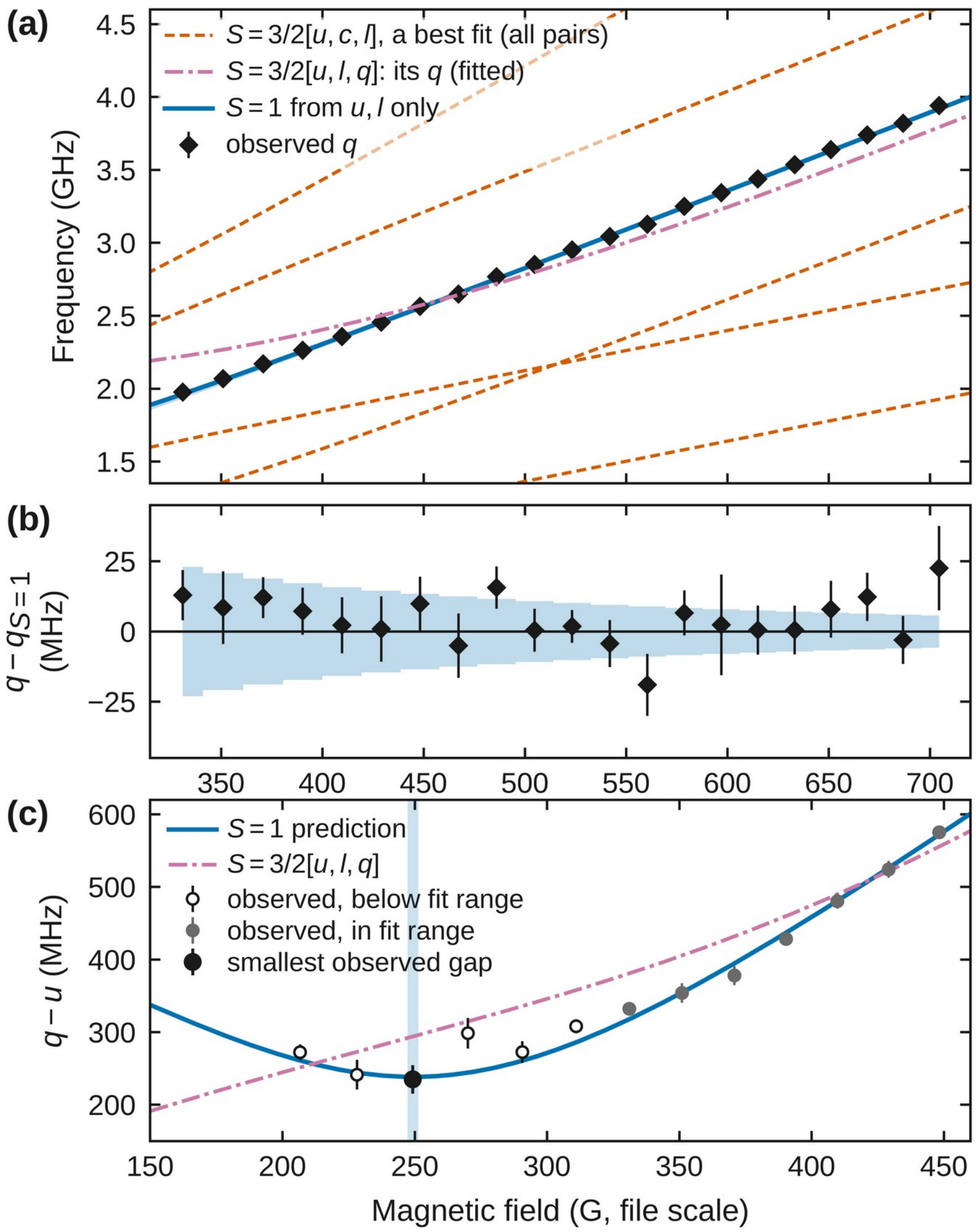

**FIG. 3.** (a) Positions of $q$ in defect 2 (diamonds), compared with the $S=1$ prediction fitted to $u$ and $l$ alone (line; band gives the 95% bootstrap interval), all transitions of one best $S=3/2\left[u,c,l\right]$ fit (dashed), and the predicted $q$ of the $S=3/2\left[u,l,q\right]$ fit with free $g$ (dash-dotted). (b) Residual of $q$ relative to the $S=1$ prediction, with $\pm1.96$ bootstrap standard deviations and no window correction. (c) Separation $q-u$ for the joint $S=1$ fit (line; shading marks $B^{*}\pm1\sigma$). Open and grey symbols mark columns below and within the fit range, respectively. The filled symbol marks the smallest observed separation under relaxed detection criteria. The $S=3/2\left[u,l,q\right]$ fit (dash-dotted) has no minimum.

### D. A strongly rhombic spin-1

A joint fit of the $S=1$ + $c$ model to 166 points gives $\chi^2=236$, with $D'=0.801\pm0.006\pm0.006$ GHz and $\left|E\right|=0.291\pm0.009\pm0.006$ GHz in the frame of Eq. (2). The first uncertainty is statistical and the second systematic. In the conventional frame, $E/D=0.30\pm0.01$. Measurements along one field direction constrain the orientation only through $n\cdot D\cdot n$, where $n$ is the field direction and $D$ is the zero-field tensor [14]. The fit is poor at the stated errors: $\chi^2/dof=1.28$ for the 108 positions of $u$, $l$ and $q$. Bringing this value to 1 requires a 6.5 MHz error floor [14]. When fitted to $u$ and $l$ alone, the $S=1$ model predicts $q$ with an rms error of $10\pm4$ MHz without window correction [Fig. 3(a,b)]. This prediction restates Eq. (1).

The avoided crossing offers a weak retrospective check. Near $B_c$, the field tilt mixes the two lower levels, producing a minimum in $q-u$. Fitted to columns at $B \geq 331$ G, the $S=1$ model predicts this minimum at $B^* = 249 \pm 2$ G, with a gap of $238 \pm 11$ MHz. Below the fit range, the smallest observed $q-u$ is $235 \pm 20$ MHz at 249 G [Fig. 3(c)], using relaxed detection criteria [14]. The axial $S=3/2$ model also has a level crossing, at 263 G.

### E. Testing an $S=3/2$ model for $u$, $l$ and $q$

The field dependence of the lines allows us to test the alternative left open by the combination rule. We fit the $S=3/2[u,l,q]$ model with $c$ separate, allowing $D$, $E$, the field direction and the $g$ tensor to vary, and testing every assignment of lines to level pairs. We used the same 108 line positions as for the $S=1$ fit and cross-checked the fitted results [14]. The tested fits do not reproduce both the low-field curvature of $q$ and the nearly linear high-field behavior of $u$ and $l$ with $g$ near 2. The best $S=3/2[u,l,q]$ fit requires an effective $g$ of 2.32 along the field. It still misses $u$, $l$ and $q$ by 53, 62 and 118 MHz rms, respectively, compared with 9.8, 8.1 and 8.5 MHz for $S=1$. In the 6 unfitted columns at 207–311 G, the two best $S=3/2[u,l,q]$ fits miss $u$ and $q$ by 393–518 MHz rms. Their predicted $q-u$, equal to $l$, has no minimum [Fig. 3(c)]. By comparison, the $S=1$ model predicts the two lines to 10 and 19 MHz.

Across all 200 column-bootstrap resamples, the $S=3/2[u,l,q]$ model gives at least 27 times the $\chi^2$ of the $S=1$ model. This result persists when $g$ is restricted to 1.8–2.2, a lower-window offset is included, 6 re-extractions are used, or the fit is restricted to the 331–705 G columns [14]. The difference, $\Delta\chi^2 = 7942$ ($\chi^2 = 8076$ compared with 133.2), rejects this particular $S=3/2[u,l,q]$ assignment. It does not establish a unique interpretation: maps simulated from other alternatives reach this difference in 90–97% of cases, and 25% of the difference comes from the 3 points on $q$ at 331–371 G [14]. The line-position fits decisively reject this assignment, but the full spectral maps give weaker evidence when residuals are correlated along the field: $z = 1.1$–$1.9$ for the best assignment and 2.1–6.0 in Ritz form [14]. These comparisons select among the tested models; a larger manifold can still contain three levels with the same connections.

Among the surviving interpretations, the $S=1$ model must also account for the amplitude of $q$, whereas the ladder treats $q$ as an independent line. Linear response predicts $A_q / A_u = 0.06$–$0.11$, well below the observed 0.76, and does not reproduce the field dependence. Optical-cycle models can reproduce both the ratio and its flat field dependence, either with an isolated $S=1$ under a polarized drive or with an $S=1$ whose signal is read out through a coupled spin-1/2. Test (v) probes these fitted models. So far, no physically consistent fit has reproduced the line heights, fluorescence transients and photon bunching using shared intrinsic rates [14]. The absence of the predicted fourth-level line in the $S=3/2[u,l,q]$ model does not distinguish the interpretations [14].

### F. The central line

The cw spectra do not fix the microscopic origin of $c$. It is 2.9 times wider than $u$, while hyperfine coupling consistent with the linewidth limit gives a ratio of at most 1.1 [14]. Pulsed ODMR [1] was recorded at a field not specified in the data files, estimated as about 994 G from $c$. It shows $c$, $u$ and a fitted $l$ line at its allowed width bound. The latter is $-39 \pm 14$ MHz from its cw track, whereas $u$ agrees with that track. The line $q$ falls outside the recorded window [14]. Thus, both $c$ and the manifold containing $u$ persist in the dark. The only microwave-prepared transient for defect 2 was inferred to drive $c$, the line with the largest contrast [14].

Several mechanisms could produce $c$. The line $c$ may belong to the $u$, $c$, $l$ ladder, with $q$ separate, as discussed above. Alternatively, it may arise from a second spin coexisting with the $S=1$, with or without coupling; from another charge configuration that alternates with it, as in the charge-transfer model [8,12]; or from a second center in the focus. The $u$, $c$, $l$ pattern recurs in every group-II emitter [1], as the ladder predicts. This recurrence makes an accidental neighboring emitter unlikely, although antibunching alone does not exclude an emitter that is systematically paired with another center.

### G. Limitations

The evidence rests on defect 2. Defect 3 is the only other group-II emitter with a detected $q$ line, but there $u+l\approx 2c$, so the line positions do not distinguish the interpretations. Its $q-2c$ also differs from that of defect 2 at the same fields [14]. Defect 4 provides no additional evidence, and none of the models fits the angle scans [14]. The closure increases with field by $+25.6\pm 9.9$ MHz per 100 G, a drift that our noise model does not capture: none of the simulated $S=1$ maps reaches it [14]. Finally, our $S=3/2$ tests are restricted to second-order spin Hamiltonians. We have not tested field-linear terms cubic in $S$, or models with $S=2$ or $S\geq 5/2$.

## V. Proposed tests

### A. Zero field

At zero field, the three $S=1$ lines occur at $D'-|E|$, $2|E|$ and $D'+|E|$. Across all fits, their frequencies span $0.50$–$0.54$, $0.56$–$0.60$ and $1.09$–$1.11$ GHz, respectively; these ranges do not account for the window shift of $l$. The frequencies again satisfy $f_1+f_2=f_3$ [Fig. 4]. Their shifts are quadratic and even under $\pm B$, reaching at most $3.1$ MHz at $1$ mT and $27.0$ MHz at $3$ mT. The two lower-frequency lines may merge [14]. For the fitted ladder at the recorded field direction, $u, l\to 0$, while $c$ approaches the Kramers line near $0.97$–$1.00$ GHz, whose components split linearly [14].

For an isolated $S=1$, the branches $u$ and $l$ must approach these finite frequencies with an even, quadratic shift when tracked continuously from high field; the matching criterion is given in Ref. [14]. If the branches instead approach zero or split linearly, this interpretation is falsified. The line $q$ provides an additional check. A coupled partner would introduce a much smaller linear term [14]. Missing zero-field contrast alone would not falsify the model, because the contrast can vanish at low field, as observed for group I [1] and in hBN [12]. The test therefore requires branches that can be observed and tracked.

### B. Experimental priorities

We propose the following tests, in order of priority. (i) Sweep defect 2 from $-3$ to $+3$ mT and back under a single field history, using low microwave power and a filtered drive [14]. (ii) Record $u$, $l$ and $q$ in one frequency window, or in interleaved windows, to remove the window shift of $l$. This measurement can distinguish the $S=1$ interpretation from a ladder accompanied by an independent line. (iii) Repeat test (i) on a second group-II emitter with $u+l\neq 2c$. (iv) Use two-tone ODMR: driving $u$ should change the contrast of $l$ and $q$, whereas in the ladder it changes $c$. A line that shares no level with $u$, such as $c$ or the independent $q$ in the ladder interpretation, remains unaffected only if its spin and optical cycle are dynamically independent of those of $u$. Spin coupling or charge transfer can also change its contrast [14]. (v) Measure the contrast and Rabi frequency of every line as a function of microwave power at one field. At $+6$ dB, the best isolated-$S=1$ fits predict a decreasing ratio with $q/u\leq 0.56$, while the coupled-spin fit predicts an increase to $1.70$ [14]. (vi) Compare Rabi frequencies at a common microwave frequency by tuning $B$. (vii) Record the zero-phonon line with each ODMR line. (viii) Finally, acquire a dense, low-power map around $B^*$ [14].

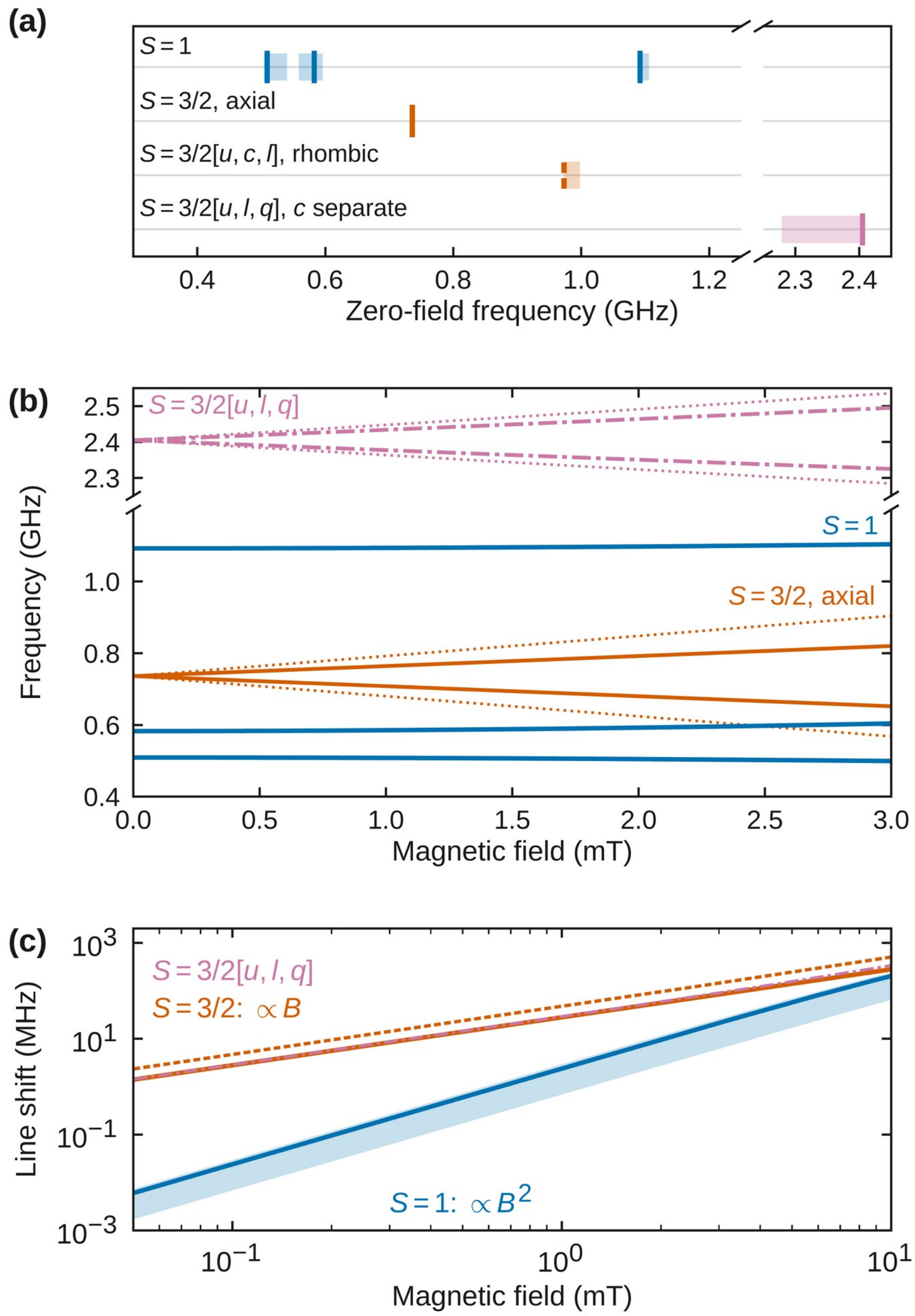

**FIG. 4.** Zero- and weak-field predictions of the tested spin Hamiltonians. (a) Zero-field frequencies for $S=1$ (bands spanning all fits), axial $S=3/2$ with the parameters of Luo *et al.* [1], rhombic $S=3/2[u,c,l]$ (band spanning the two best fits) and, beyond the axis break, $S=3/2[u,l,q]$ (band spanning fits with $g$ restricted to 1.8–2.2 through fits with $g$ free; line shows the fit with $g$ free). (b) Field dependence for $S=1$ at the measured field direction, axial $S=3/2$ along its axis (allowed lines solid; $\Delta m_s=2$ lines dotted) and, above the break, the $S=3/2[u,l,q]$ components (strong components dash-dotted; weak components dotted). (c) Largest frequency shift; an $S=3/2$ line spreads by twice this amount. Curves show $S=1$, axial (solid) and rhombic (dashed) $S=3/2[u,c,l]$, and the strong $S=3/2[u,l,q]$ components (dash-dotted). The band spans $S=1$ predictions with $B$ along each principal axis. All curves use $g$ on the field scale of the data.

## VI. Microscopic origin

The large $E$ excludes $C_{3v}$ symmetry. The microscopic defect is not yet identified. We screened Ga-vacancy defects with a hybrid functional ($\alpha=0.31$) in a 92-site cell, including only spin–spin contributions and testing a subset of charge states. Every computed localized two-hole $S=1$ state gives $|D|=4.8$–$5.2$ GHz, or $5.8$–$6.3$ times the measured $|D|$. Even the classical dipolar contribution is too large and too axial [14], and the screen does not support averaging through hole hopping. For four Fe complexes, $M_S=1$ lies $1.0$–$1.4$ eV above $M_S=2$ (we did not test $S=2$ spin Hamiltonians against the ODMR tracks). $Co_{Ga}^-$ has been proposed as an $S=3/2$ emitter candidate [15]. Assuming one center, three distinct single-photon electron transitions and a linear Zeeman interaction, the observed tracks require a zero-field gap of $\Delta_0 \leq 8.8$ GHz for $|g^T \hat{n}| \leq 2.4$ on the recorded field scale [14]. Measured $d^7$ analogues suggest larger splittings [16,17]. Embedded-cluster calculations give gaps of at least $11.9$ GHz, but they omit spin–spin coupling, are not converged with cluster size and give a smaller gap in the larger cluster [14]. These comparisons disfavor this candidate without excluding it. For NiGa−, a non-magnetic spin–orbit ground level at the static geometry disfavors the assignment [14].

## VII. Conclusions and prospects

The unexplained line lies within a few tens of MHz of the sum of the two outer lines, as expected if the three transitions connect a common set of three levels. None of the tested second-order $S=3/2$ fits containing that line reproduces the observed positions, and neither does the minimal model of Luo *et al.* A rhombic spin-1 provides an economical realization of the three-level interpretation.

The main open questions are experimental. Because the window shift of $l$ is only measured near 950 G, the line positions alone rule out an $S=3/2$ ladder with an accidental $u+l$ line only if that shift is constant across the sweep; otherwise the case for spin-1 rests on parsimony. The available cw and transient traces cannot distinguish between an isolated $S=1$ under a polarized microwave drive and an $S=1$ read out through a coupled spin-1/2, nor do they explain why $q-(u+l)$ drifts with field.

A zero-field measurement offers the next direct test. Observing the three predicted lines with quadratic field shifts would identify a strongly rhombic, non-Kramers three-level manifold in defect 2, most simply realized by a spin-1. Test (iii) would examine whether this interpretation also applies to group II. If confirmed, the zero-field $S=1$ lines would be clock transitions [18–20]: insensitive to magnetic fields to first order, while remaining sensitive to strain, electric fields and nuclear spins [14,21,22].

## Acknowledgments

We thank the authors of Ref. [1] for making their raw data and analysis code publicly available [9]. We use these materials for non-commercial research under the CC BY-NC 4.0 license. This work used the Anvil supercomputer at Purdue University through allocation MAT260089 from the Advanced Cyberinfrastructure Coordination Ecosystem: Services & Support (ACCESS) program, which is supported by U.S. National Science Foundation grants 2138259, 2138286, 2138307, 2137603, and 2138296 [23,24]. This research also used resources of the Oak Ridge Leadership Computing Facility at Oak Ridge National Laboratory, which is supported by the Office of Science of the U.S. Department of Energy under Contract No. DE-AC05-00OR22725.

## Data availability

The raw data are available in Ref. [9] (doi:10.7298/azf0-hc03). The line tables, the electronic-structure outputs and a table identifying the source file for each reported numerical result are archived at Zenodo (doi:10.5281/zenodo.23744355). The analysis code, computational environment, random seeds, extraction scripts and electronic-structure inputs are available from the author on request.

# Supplemental Material for “A rhombic spin-1 interpretation of a room-temperature GaN emitter”

Daniel Hashemi
*Department of Physics, Optical Engineering and Nanoengineering, Rose-Hulman Institute of Technology, Terre Haute, Indiana 47803, USA*
hashemi@rose-hulman.edu

Equation and figure numbers without an “S” prefix refer to the main text. References in Part A use [1]–[23]; references in Part B use [B1]–[B24]. Sections marked “R” are in Part B of this document, which gives the detailed methods and checks.

## S1. Data and line extraction

### Data and conventions

We analyze the public data and scripts of Luo *et al.* [1]. Defect 2 was mapped in two separate frequency sweeps, 0.01–1.99 and 1.80–4.00 GHz, at the same field setpoints (92–1338 G). The archive contains no acquisition times, sweep order or magnet log, and no zero-field or power-dependent measurements. Section R20 gives the full inventory.

*Field scale.* Experimental fields are quoted on the gauss scale recorded in each data set (“file G”); model fields are in mT. Luo *et al.* calibrated the field against magnet position, using $g=2$ for the most visible lines [2]. In the archived files, the central line of defect 2 disperses with $g_{file}=1.956\pm0.003$. Neither this value nor those for defects 3 and 4 is a calibrated $g$ factor: the scale differs between defects, and the defect-3 intercept remains unexplained (Sec. R1). We therefore base the main argument on combinations of line positions within the same column, which do not depend on the field scale. These combinations do, however, assume that both windows of a column were measured at the same field (Sec. S2).

*Frame.* The parameters $D'$ and $E$ in Eq. (2) are defined in the fit frame, whose $z$ axis follows the $P_1$ axis of the traceless principal values $P_1>P_2>P_3$. Throughout, “$B \parallel z$” refers to this axis, rather than $c$. Because $|E|/D'>1/3$, the fit frame differs from the conventional (Bleaney) frame, whose $z$ axis is the axis of largest $|D_{ii}|$, giving $|E|\le|D|/3$. Both frames describe the same Hamiltonian and frequencies; we report the parameters in each.

We do not call $c$ an $S=1/2$ spin. In the phenomenological $S=1+c$ model (Sec. S3), the line may arise from a coexisting spin, another charge configuration of the same defect, or a neighboring center.

### Line extraction

*Column model.* We fit each field column with a sum of spectral lines and a polynomial baseline, allowing a separate noise level for each window. The primary model uses a pseudo-Voigt profile for $c$, Gaussian profiles for $u$, $l$ and $q$, and a linear baseline. The replicas at $c/2$ and $u/2$ ($g=1$), identified by Luo *et al.* as amplifier artifacts [2], have positions and widths tied to their parent lines, with free, non-negative amplitudes. We take half the range across line-shape and window variants as the systematic error. Section S2 describes a broader test of line shapes and replicas.

*Tracking.* We first fit $c$ and its replica alone in each column. The residual map then guides the tracks of $u$, $l$ and $q$ outward from their strongest columns, after which we refine the joint fit iteratively in each column. For $q$, the tracking coordinate is the offset from $2c$ (Sec. R2); its extrapolation follows the fitted $q$ track, with the offset free to vary. The search imposes neither a zero offset nor the combination rule.

*Detection and errors.* We classify a line as detected if removing it increases $\chi^2$ by at least $16$, its amplitude exceeds $3$ standard errors, and its position error is below $30$ MHz. Recovering four-line columns in defect 3 requires a relaxed criterion ($\Delta\chi^2 \geq 9$, with blended $l$ retained; Sec. S2). Position errors come from the Jacobian covariance, scaled by $max\left(1, \chi^2_{red}\right)$. Columns in which $l$ lies within $80$ MHz of the $c/2$ or $u/2$ replica are classified as blended and excluded from the means (Table S1; Sec. R1 for sample and excitation; Secs. R2 and R13).

## S2. The combination rule in the data

### The faint branch and its curvature

Luo *et al.* already reported the $g \approx 4$ branch [2]. We therefore test how reliably it can be recovered from the archived data. In defect 2, the branch appears in all 21 columns from 331 to 705 G, with a track statistic of $T = 26.1$. This is a matched-filter statistic, not a Gaussian standard deviation. We applied the same search to 4000 null maps that retain the strong lines, replicas, baselines and correlated noise but omit $q$. The null statistic has a mean of 2.5 and a maximum of 4.4. Across the 27 residual-map statistics, the smallest observed value is 24.8 and the largest null value is 7.3, so on these statistics no null map reaches one third of the data value. Defect 4 does not separate from its null. This comparison shows that the track is not an artifact of the search, although the spin assignment does not depend on this check.

The quadratic term fitted to the $q$ positions is $\left(8.2 \pm 1.4\right) \times 10^{-4}$ MHz/G$^2$. It differs from zero by $6.1\sigma$ against the nominal field, by $4.6\sigma$ when $c$ serves as the field ruler (assuming $c$ is linear in $B$), and by $3.4\sigma$ after including a $5$ MHz error floor and a cubic field distortion shared by all lines. This curvature is consistent with the $S=1$ form $f_q = 2\sqrt{E^2 + \left(g\mu_B B\right)^2}$. It also allows an $S = 3/2\left[u, c, l\right]$ ladder accompanied by a separate curved line at $q$ (Secs. R5 and R13).

### The closure and the window shift of $l$

*Uncorrected value.* For the 8 columns at 560–705 G in which all four lines are detected and $l$ is unblended (Table S1), we obtain $q - \left(u + l\right) = +4.4 \pm 5.9\left(stat\right) \pm 4.1\left(sys\right)$ MHz. This residual is small compared with the 60–100 MHz linewidths and depends on the separation cut. Across 36 supported line-shape and replica variants, it ranges from $+3.0$ to $+6.3$ MHz. Allowing a free asymmetry for every line expands the range to $-4.5$ to $+18.4$ MHz, but the data reject that model. Among the extraction choices, only the replica model pulls $q$ toward $u + l$: omitting replicas gives $+9.1$ rather than $+4.4$ MHz. The data require the replicas.

**TABLE S1.** Same-column combinations in MHz for defect-2 columns in which all four lines are detected. Uncertainties are $\pm 1\sigma$ from the line fits. The 505 G column, where $l$ lies within 80 MHz of $c/2$, is excluded from all means. Section R3 gives the dependence on this cut.

| $B$ (G) | $q - \left(u + l\right)$ | $q - 2c$ | $q - \left(u + c\right)$ | $q - \left(c + l\right)$ |
|---|---|---|---|---|
| 505* | $26 \pm 9$ | $95 \pm 8$ | $-667 \pm 9$ | $788 \pm 7$ |
| 560 | $-25 \pm 13$ | $58 \pm 12$ | $-699 \pm 12$ | $731 \pm 12$ |
| 579 | $-2 \pm 9$ | $90 \pm 8$ | $-675 \pm 9$ | $762 \pm 8$ |
| 615 | $-12 \pm 16$ | $89 \pm 9$ | $-686 \pm 10$ | $763 \pm 15$ |

| $B$ (G) | $q-(u+l)$ | $q-2c$ | $q-(u+c)$ | $q-(c+l)$ |
|---|---|---|---|---|
| 633 | $17\pm11$ | $77\pm9$ | $-681\pm9$ | $775\pm10$ |
| 651 | $6\pm13$ | $82\pm10$ | $-680\pm10$ | $767\pm13$ |
| 669 | $27\pm14$ | $83\pm9$ | $-671\pm9$ | $781\pm13$ |
| 687 | $10\pm15$ | $67\pm8$ | $-696\pm9$ | $773\pm14$ |
| 705 | $15\pm18$ | $96\pm15$ | $-680\pm15$ | $790\pm17$ |
| mean (8) | $4.4\pm5.9$ | $80\pm4$ | $-683\pm3$ | $768\pm6$ |

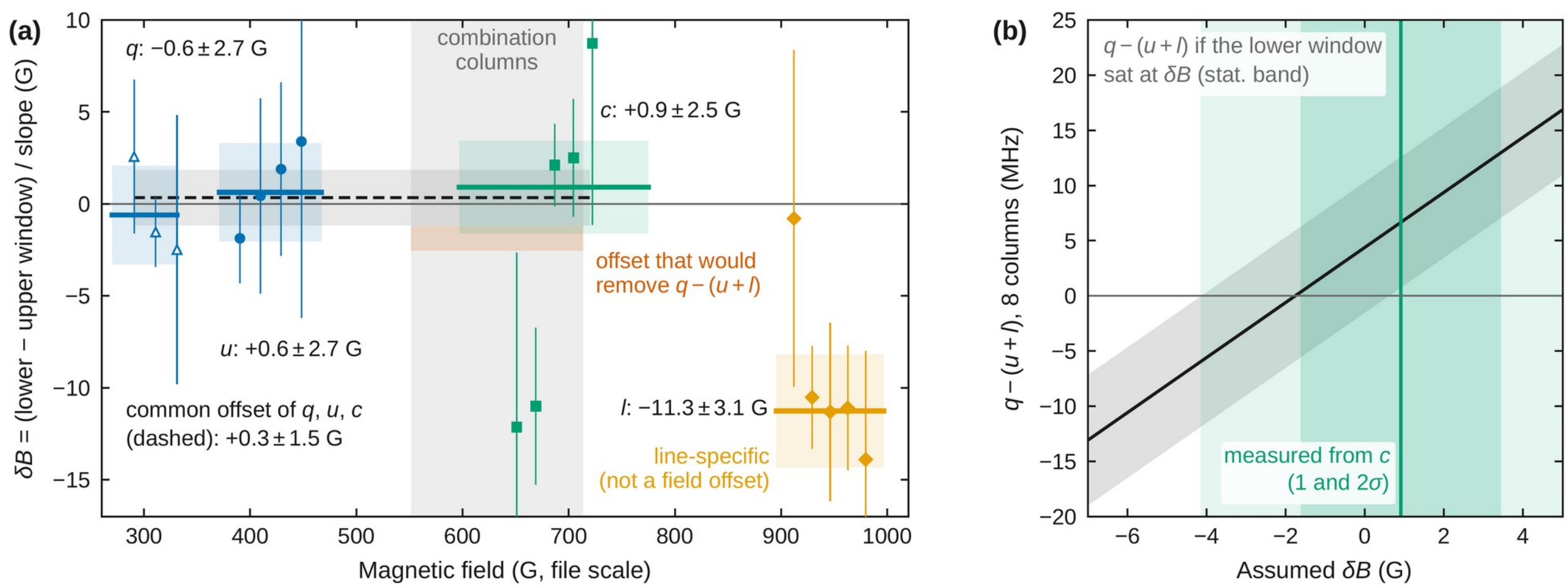

**FIG. S1.** Window offsets for defect 2. (a) For each line crossing the overlap, the lower-minus-upper frequency difference is expressed as a field offset. Points show individual columns, and bars give the joint value for each line, with a $\pm1\sigma$ band (stat$\oplus$sys). The dashed line shows the common offset inferred from $q$, $u$ and $c$. Grey marks the combination columns; the red-orange band shows the offsets that would remove $q-(u+l)$ for the supported variants. (b) $q-(u+l)$ in the combination columns as a function of $\delta B$, with a band showing the statistical error. The offset measured from $c$ is shown at $1\sigma$ and $2\sigma$.

*Window offset.* In the columns used to test the combination rule, $l$ lies in the lower window, while $q$ and $u$ lie in the upper window. If the lower window was recorded $\delta B$ above its nominal field, the measured closure is reduced by $s_l \delta B$, where $s_l = 2.5$ MHz/G is the slope of $l$. Lines crossing the 1.80–1.99 GHz overlap appear twice in a column, so dividing their frequency shifts by their slopes gives $\delta B$ (Fig. S1). The estimates from $q$, $u$ and $c$ agree on $\delta B = +0.3\pm1.5$ G, whose sign is opposite to that needed to explain the residual. We apply no offset correction and instead carry the uncertainty as a window term of $\pm6.3$ MHz. Defect 1, used only for this check, gives $-1.40\pm0.65$ G. An offset of $-1.8$ G, which would remove the residual, is not excluded.

*The shift of $l$.* At 895–997 G, where $l$ crosses the overlap, its positions in the two windows differ by $31\pm8$ MHz. Its amplitude changes by a factor of 1.33 between windows, compared with 1.01–1.11 for $c$, $u$ and $q$. In the same columns, the $u/2$ replica recorded in the lower window does not move relative to $u$, disfavoring a field offset at $2.9\sigma$. The unexplained shift is therefore specific to $l$; Sec. S5 proposes tests. In the combination columns, $l$ has no fiducial, so a shift of this size cannot be bounded. It is also not covered by the window term, which allows only for a common field offset. The archive does not identify which window carries the shift (Sec. S1).

*Carrying the shift through.* Joint fits to the line positions with explicit window models (Table S2) place $q-(u+l)$ at $2.3\sigma$ from zero if the lower window carries the shift, or at $+2.8\pm7.9$ MHz if the upper window does. With the attribution unknown, the equal mixture of these two possibilities gives Eq. (3) of the main text.

Equal weighting is a prior assumption, and both possibilities assume that the shift is constant with field. The quoted 95% interval is the central interval of this mixture, conditional on both assumptions.

**TABLE S2.** Defect-2 values of $q-(u+l)$ for the 8 columns in Table S1, obtained by fitting the line positions jointly with a window nuisance term. The upper window is the reference, and the quoted errors come from the joint fit. B0 denotes the uncorrected model. Section R20 gives other window models.

| model | window nuisance | $q-(u+l)$ (MHz) |
|---|---|---|
| B0 | none | $+4.4\pm6.9$ |
| B2f | constant shift of $l$, lower window | $-24.8\pm10.9$ |
| B2uf | constant shift of $l$, upper window | $+2.8\pm7.9$ |
| B2f/B2uf | attribution unknown (equal mixture) | $-11.1\pm16.7$; 95%: $-43$ to $+16$ |

*Drift with field.* Across the 8 combination columns, the closure increases by $+25.6\pm9.9$ MHz per 100 G, with the lowest-field column driving the trend. Neither an exact combination rule nor a constant window shift predicts such a trend, and no $S=1$ map simulated with our noise model reaches the observed drift. Thus, the $S=1+c$ model combined with our noise model does not reproduce every statistic of the data. This discrepancy does not strongly favor any tested alternative. A global fit over 331–705 G, with both offset and slope fitted, gives $-6.4\pm4.5$ MHz near 520 G. The sign of the residual is therefore not robust across field ranges.

### How far $q$ lies from $2c$

Three interpretations can place $q$ at or near $2c$. For an axial $S=3/2$ ($E=0$, with $B$ along $z$), the levels $E_m=D\left(m^2-5/4\right)+\gamma Bm$, where $\gamma=g\,\mu_B/h$, give the allowed lines above the crossing ($\gamma B>2D$): $l=\gamma B-2D$, $c=\gamma B$ and $u=\gamma B+2D$. The relation $u+l=2c$ then holds identically, and only $q-2c$ distinguishes a line at $u+l$ from one at $2c$. A separate line with twice the $g$ of $c$ also gives a zero difference. For example, a $\Delta m_s=2$ line of the spin responsible for $c$ has $q-2c=0$. A weakly coupled pair need not satisfy this relation.

*Measurement.* Across the 21 columns at 331–705 G, we find $q-2c=102\pm7$ MHz. The difference is at least 58 MHz in every column and decreases by $20\pm3$ MHz per 100 G. With symmetric line shapes, every column is at least $4.8\sigma$ from zero. Supported dispersive shapes reduce the weakest column (560 G) to 3.1–$4.2\sigma$. Allowing the rejected model with a free asymmetry for every line reduces the minimum to $2.1\sigma$, at 705 G. Across 44 variants, the mean remains between 97.8 and 108.8 MHz.

Because $c$ lies in the lower window in these columns, a field offset changes $q-2c$ by twice the slope of $c$ per gauss. Removing even the smallest column value would require an offset beyond $-10$ G, which $c$ alone excludes at $4.4\sigma$. These measurements also constrain three other interpretations. Amplifier replicas appear at $c/n$ rather than $2c$. A two-photon resonance of the same $S=1$ would place $c$ at $q/2$; testing $c-q/2=-\left(q-2c\right)/2$ is therefore the same measurement. Finally, no column lies within 573 MHz of the $S=3/2$ sums $u+c$ or $c+l$.

*Hidden spin pair.* For a weakly coupled pair with equal $g$, the $T_+\leftrightarrow T_-$ line occurs at the sum of the two $\Delta m_s=1$ lines. If both are hidden under $c$, its fitted center lies near their weighted mean. Unequal weights can therefore make $q-2c$ approach their splitting, which the width of $c$ could conceal (Sec. S4). Extrapolating the trend of $q-2c$ gives $+204\pm18$ MHz at $B=0$. This excludes only a pair whose splitting comes from a $g$ difference. A centroid bound disfavors the pair: the $T_+\leftrightarrow T_-$ line lies within 5.0 MHz of twice the strength-weighted centroid of the $\Delta m_s=1$ lines, and the best pair fit to $c$ and $q$ gives $\chi^2=2451$ on 79 points. The bound does not apply when the contrast weights differ from the transition strengths, so the pair is not excluded. Test (iv) in Sec. S5 checks the hidden-pair interpretation directly.

For hBN, the preprint of Whitefield *et al.* assigned a $g=4$ line to the $S=1$ double-quantum transition [3], whereas their published Supplement leans to the $T_+ \leftrightarrow T_-$ line of a weakly coupled pair [4]. Their spectra cannot distinguish these assignments. An exact $S=1$ calculation places the line only 5.4 and 8.5 MHz above $2c$ at the two measured fields (field along $z$), while an uncoupled pair places it at $2c$. In GaN, $q$ lies 58–168 MHz from $2c$ (Secs. R3, R12 and R15).

#### Defect 3, and what stays open

Under the relaxed detection criterion, only 3 columns of defect 3, at 630–690 G, show all four lines. They give $q-(u+l)=14\pm17$ (sys 28) MHz and $q-2c=-7\pm23$ (sys 10) MHz. Because $u+l\approx2c$, these positions cannot distinguish the $S=1$ interpretation from a line at $2c$. The value of $q-2c$ also differs from the 67–83 MHz found for defect 2 at the same fields. Either $c$ is not rigidly tied to the $S=1$, or the $S=1$ of defect 3 has different parameters or orientation. A spin-Hamiltonian fit based on these 3 columns with relaxed detection would not resolve the question, so we have not attempted one.

For defect 2, the shift of $l$ in the combination columns remains unbounded, and the closure drifts with field (Sec. S2). Drive-induced shifts were not measured. Test (ii) in Sec. S5 would remove the anomaly of $l$ (Secs. R3, R12 and R20).

## S3. Spin-Hamiltonian fits

#### The $S=1+c$ joint fit

In the $S=1+c$ model, an $S=1$ accounts for $u$, $l$ and $q$, while $c$ is treated as a separate line that is linear in $B$ with zero intercept. We assign each position a variance of $\sigma_{fit}^2+(5\,MHz)^2$, where $\sigma_{fit}$ is the line-fit error. Statistical errors are standard deviations from a column bootstrap; systematic errors are half the range across field-range and line-selection variants. Fitting at $B\geq331$ G, with $q$ included at 331–705 G (166 points, $\chi^2=236$), the $S=1+c$ model gives $D'=0.801\pm0.006\pm0.006$ GHz and $|E|=0.291\pm0.009\pm0.006$ GHz (stat, fit range) in the frame of Eq. (2). In the conventional frame, $E/D=0.30\pm0.01$. The traceless principal values are $P_1=+0.53$, $P_2=+0.02$ and $P_3=-0.56$ GHz, with undetermined overall sign. For the 108 positions of $u$, $l$ and $q$ in Sec. S3, the fit remains poor relative to the stated errors: $\chi^2/dof=1.28$. A per-point error floor of 6.5 MHz brings this ratio to 1.

A straight line describes $c$ over 331–705 G ($\chi^2=25$ for 19 degrees of freedom), consistent with its use as the field ruler in Sec. S2. It does not describe the full 311–1322 G range. Either $c$ is nonlinear in $B$ over that range or the field scale becomes nonlinear at high field. Same-column combinations are unaffected. The joint fit extends to 1322 G, and its errors include the systematic uncertainty from the fit range.

A measurement along one field direction constrains the orientation only through $n\cdot D\cdot n=0.491\pm0.001\pm0.002$ GHz, where $n$ is the field direction in the tensor frame. For isotropic $g$, the characteristic polynomial of Eq. (2) depends on $n$ only through this quantity. The polar angle $\theta$ and azimuth $\phi$ of $n$ can therefore vary together along a one-parameter family with identical spectra, giving $\theta=11$–$17^\circ$ from the $P_1$ axis. Combining the $C_{1h}$ site symmetry with the approximate field direction of Luo *et al.* [2] reduces this family to discrete tilts of the $P_1$ axis from $c$. These estimates provide targets for the calculations (Sec. R4).

### The $u$, $c$, $l$ ladder and the line $q$

*Ladder fits.* Fitting the ladder of Luo *et al.* to $u$, $c$ and $l$ with free $D$, $E$, $\theta$, $\phi$, $g_{\parallel}$ and $g_{\perp}$ gives several fit modes. Only the zero-field invariant $2\sqrt{D^2+3E^2}=0.97\pm0.02$ GHz is stable. Even the best fit misses $q$ by at least $321$ MHz in every column ($609$ MHz rms), including two- and three-photon resonances. Assigning all four lines to one $S=3/2$ gives a best fit, out of $360$ assignments, that requires an effective $g$ of $|g\cdot n|=1.45$. It yields $\chi^2=2902$ on $71$ points at $331$–$705$ G, compared with $41$ for the $S=1+c$ model. At the precision of this extraction, the axial $S=3/2$ with the published parameters also misses the strong lines ($\chi^2=1413$ on $145$ positions, compared with $221$ for the $S=1+c$ model).

*An independent line at $q$.* A calibrated test (Table S3) compares a ladder accompanied by an empirical quadratic line at $q$ (H0) with the $S=1+c$ model, which predicts $q$ (H1). The statistic $D=\chi^2(H0)-\chi^2(H1)$ uses the positions of all four lines together with the window-overlap shifts. Through $q$, it also includes the closure information. Replicates from each hypothesis use the same window data model, with its $\pm47.2$ MHz prior box, and a smooth field-scale distortion common to all lines; they add Gaussian noise to positions, not to maps, and both models are refitted to each.

Assuming a constant shift of $l$ in either window (Sec. S2), H0 gives $p\leq0.0075$ (the largest estimate across windows and error models), while H1 gives $p\geq0.04$ and is marginal in one case ($p=0.042$). The constant-shift assumption is neither established by measurement nor supported by the closure drift. No significance level was fixed in advance; at 5%, the $p=0.042$ case is a nominal rejection of $H_1$ that the $n=400$ sample does not resolve. These $p$ values describe tails of $D$; they are not goodness-of-fit tests. Allowing the shift to vary linearly with field puts the observation midway between the predictions and leaves the test inconclusive; with per-line errors, the data then mildly favor the ladder, with $\ln\mathrm{LR}=-1.0$. The positions of $u$ and $c$ alone disfavor the ladder only weakly ($p=0.010$–$0.075$ across noise models).

**TABLE S3.** Calibrated comparison of H0 and H1 (Sec. S3). Window models are B2f and B2uf from Table S2, or B3f, in which the lower-window shift of $l$ is linear in $B$. Errors are either as fitted ("paper") or inflated separately for each line. $D_{obs}$ is the observed value of $D$. We use $p=(k+1)/(n+1)$, where $k$ of $n=400$ replicates from each hypothesis are at least as extreme as the observation. $\ln$LR is the Gaussian log-likelihood ratio H1/H0.

| window | errors | $D_{obs}$ | $p_{H0}$ | $p_{H1}$ | $\ln$LR |
|---|---|---|---|---|---|
| B2f | paper | $+16.0$ | 0.0025 | 0.085 | $+8.4$ |
| B2f | per line | $+9.0$ | 0.0025 | 0.042 | $+4.4$ |
| B2uf | paper | $+8.0$ | 0.0025 | 0.11 | $+3.5$ |
| B2uf | per line | $+3.0$ | 0.0075 | 0.12 | $+2.2$ |
| B3f | paper | $-0.3$ | 0.045 | 0.050 | $+0.1$ |
| B3f | per line | $-4.1$ | 0.17 | 0.042 | $-1.0$ |

*The coincidence premise.* The remaining reasons to disfavor the ladder are parsimony and the premise that an independent line is unlikely to follow $u+l$. Within the extraction search box, a free track satisfies the closure condition by chance with probability $1.1\times10^{-2}$, rising to $2.2\times10^{-2}$ in a narrow box around $g\approx4$. For the $S=1$ prediction of $q$ over a free track, this conditional, uncorrected comparison gives a Bayes factor ranging from $10^{4.0}$ to $10^{5.1}$ across the tested error models and search boxes. This Bayes factor uses positions without a window correction, treats the $S=1$ prediction as exact and was not recomputed with a window shift of $l$. It is not an evidence ratio for the instrument-aware hypotheses because the unresolved window shift has not been marginalized. The chance probabilities at the precision of Eq. (3) illustrate the assumed coincidence prior; they do not resolve the window-shift ambiguity. All of these numbers depend on the prior box. They quantify the coincidence premise without testing its validity (Secs. R4 and R20).

### Prediction of $q$ and the avoided crossing

When fitted to $u$ and $l$ alone, the $S=1$ model predicts $q=E_2-E_0$ with an rms error of $10\pm4$ MHz over 21 columns, without a window correction. Because $E_2-E_0=(E_2-E_1)+(E_1-E_0)$ is an identity, this prediction follows from the combination rule and the quality of the $u$, $l$ fit. It provides no independent evidence.

The avoided crossing provides a weak retrospective check. The line formulas below Eq. (2) place the crossing of $T_-$ and $T_0$ for $B \parallel z$ at $s=D'$. For the joint-fit parameters, this occurs at $B_c \approx 27$ mT (272 file G). Since $f_l=|D'-s|$, the same three levels give $q=u+l$ above $B_c$ and $q=u-l$ below it. The tilted field in the data makes this an avoided crossing: $l$ reaches a minimum where $q$ and $u$ are closest. The columns at $B \geq 331$ G predict this minimum at $B^*=249\pm2$ G, with a gap of $238\pm11$ MHz. Below the fit range, the smallest measured $q-u$ is $235\pm20$ MHz at 249 G. That result comes from a single column excluded from the fit, in which $q$ and $u$ meet only the relaxed detection criterion. The axial $S=3/2$ model with $D=368$ MHz also has a level crossing, at 263 G, so the location alone cannot distinguish the models (Sec. R6).

### An $S=3/2$ hosting $u$, $l$ and $q$

Any three levels within an $S=3/2$ quartet also obey the combination rule. We fitted this $S=3/2[u,l,q]$ model, treating $c$ separately, with $H=D\left[S_z^2-5/4\right]+E\left(S_x^2-S_y^2\right)+\mu_B B\cdot g\cdot S$, an axial $g$ tensor about $z$ and a free field direction. Because any $g$ tensor enters the line positions only through $g\cdot n$, this parameterization covers every $g$ tensor for the positions, though not for the transition strengths. We evaluated every assignment of $u$, $l$ and $q$ to level pairs using the same 108 points as the $S=1$ fit ($q$ at 331–705 G and unblended $l$). Data at one field direction do not identify all parameters of this $S=3/2$ model, so we compare residuals rather than an information criterion.

**TABLE S4.** Comparison of $S=1$ $[u,l,q]$ with $S=3/2[u,l,q]$, treating $c$ separately, on identical points (108 unless stated). The $g$ column gives bounds on the tensor components; widening them beyond 1.6–2.4 leaves the optimum unchanged. "Ritz" denotes the best assignment with $q=u+l$, "any" the best of all assignments, and "nearest" the assignment of each point to the nearest $S=3/2$ transition. Rms errors are in MHz; $\Delta\chi^2$ is measured relative to the $S=1$ fit.

| model | $g$ | assignment | rms $u/l/q$ | $\chi^2$ | $\Delta\chi^2$ |
|---|---|---|---|---|---|
| $S=1$ | free | Ritz | 9.8/8.1/8.5 | 133.2 | – |
| $S=3/2$ | 1.6–2.4 | Ritz | 53/62/118 | 8076 | 7942 |
| $S=3/2$ | 1.6–2.4 | nearest | | 5760 | 5626 |
| $S=3/2$ | 1.8–2.2 | Ritz | 99/53/107 | 13085 | 12951 |
| $S=3/2$ | 1.8–2.2 | any | | 10337 | 10204 |
| $S=3/2$ | 1.8–2.2 | nearest | | 7316 | 7183 |
| 331–705 G only (50 points) | | | | | |
| $S=1$ | free | Ritz | | 33 | – |
| $S=3/2$ | unbounded | Ritz | | 578 | |
| $S=3/2$ | 1.6–2.4 | Ritz | | 896 | |
| $S=3/2$ | 1.8–2.2 | Ritz | | 1730 | |

Within the fit range, the discrepancy is systematic. With $g$ near 2, the tested $S=3/2$ fits do not reproduce the curvature of $q$ at low field while keeping $u$ and $l$ straight at high field (Table S4, Fig. S2). The best $S=3/2[u,l,q]$ fit requires $|g\cdot n|=2.32$, well above $g=1.9560$ for the $S=1$ model and $1.9528$ for $c$. On the calibrated field scale of Luo *et al.*, however, this value of $g$ is plausible for a transition-metal $S=3/2$ center. We therefore leave $g$ free in the main comparison.

The preference persists under the checks performed. In each of $200$ column-bootstrap resamples, the $S=3/2[u,l,q]$ model has at least $27$ times the $\chi^2$ of the $S=1$ model. The result also persists when $g$ is bounded to 1.8–2.2, when a constant offset is allowed for the lower-window $l$ ($\Delta\chi^2\geq 7149$), and across $6$ extraction variants ($\Delta\chi^2\geq 7771$). It holds on the $331$–$705$ G columns alone, where the best $S=3/2$ fit requires unbounded $g$ ($|g\cdot n|=2.49$). An error floor of about $74$ MHz per position would be needed to make the model with unrestricted g acceptable on its own. This is comparable to the line widths ($u$: $92$ MHz FWHM), and even then the free-$g$ $S=3/2[u,l,q]$ model would fit worse than the $S=1$ model.

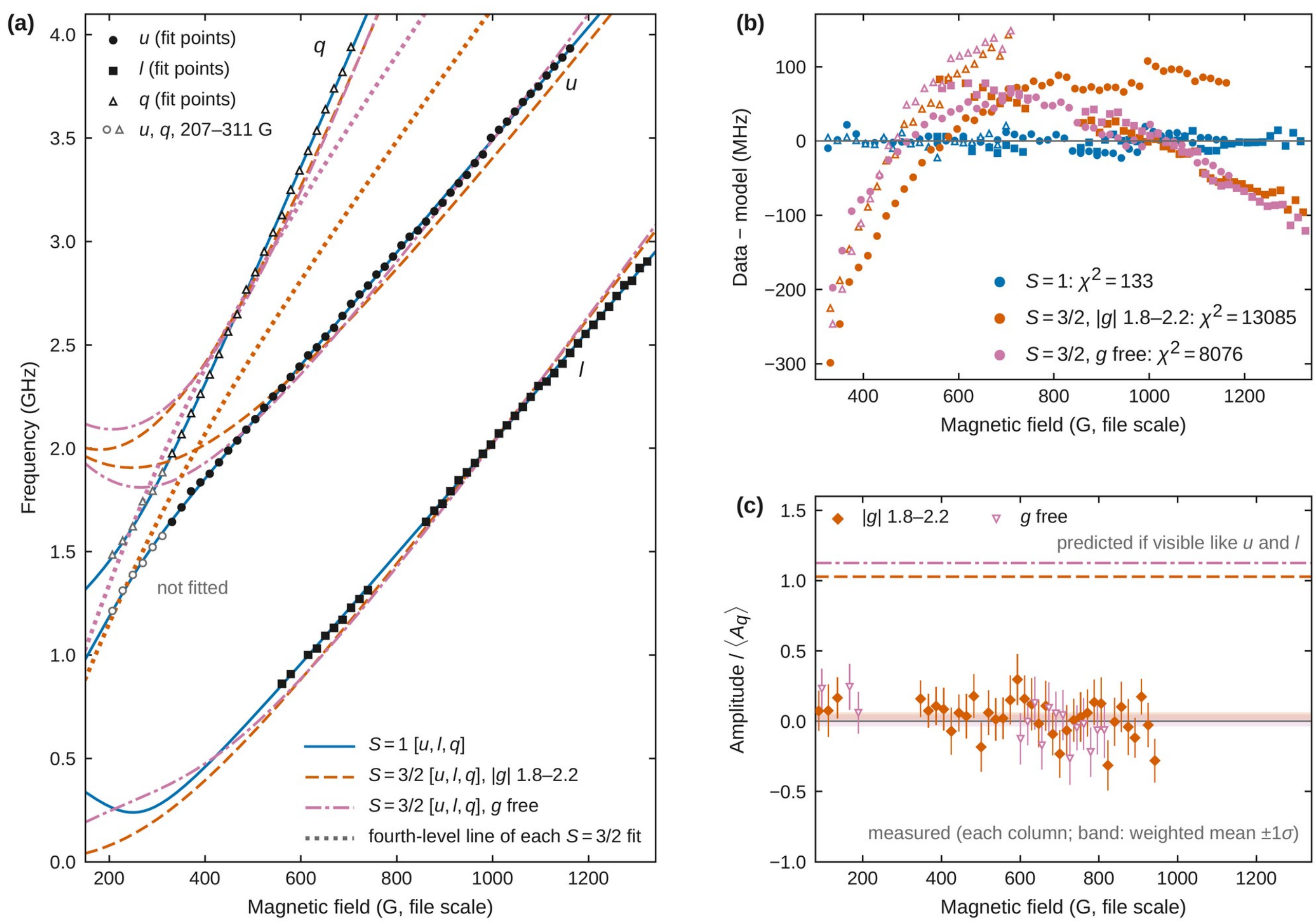

**FIG. S2.** The $S=3/2[u,l,q]$ model for defect 2. (a) Fitted positions of $u$ and $l$ (filled symbols) and $q$ (open triangles), together with $u$ and $q$ at 207–311 G, which were excluded from the fits (grey). Curves show the $S=1$ fit (solid), the best $S=3/2[u,l,q]$ fit with $|g|$ bounded to 1.8–2.2 (dashed), the fit with free $g$ (dash-dotted), and the strong transition involving the fourth level predicted by each $S=3/2$ fit (dotted). (b) Residuals on the common points: circles, $u$; squares, $l$; triangles, $q$. (c) Amplitude fitted at the predicted fourth-level line wherever it is resolved from every fitted line, in units of the mean amplitude of $q$. The band shows the weighted mean and its standard deviation. Horizontal lines show the amplitude expected if this transition were detected in the same way as $u$ and $l$.

Outside the fit range, the 6 columns at 207–311 G provide a further comparison. Here $u$ and $q$ meet the relaxed detection criterion. The $S=1$ model predicts $u$ and $q$ with rms errors of 10 and 19 MHz, respectively, whereas the two best $S=3/2[u,l,q]$ fits miss them by 393–518 MHz. In both of those fits, $q-u$ equals $l$, decreases monotonically and closes only at $B=0$. Neither therefore has an avoided crossing, although the observed minimum itself rests on a single column (Sec. S3).

The value $\Delta\chi^2=7942$ rejects this particular $S=3/2[u,l,q]$ model. Maps simulated from other alternatives reach this value in 90–97 % of realizations, and 25 % of the difference comes from the 3 $q$ points at 331–371 G. The line positions reject the model decisively, but the spectral-map comparison is less clear because residuals are correlated along the field. A column-block Newey–West statistic gives $z=1.1$–$1.9$ for the best assignment and 2.1–6.0 for the Ritz assignment. Above 705 G, the best assignment fits the maps better than the $S=1+c$ model. Each Ritz fit also predicts a fourth-level line with a linear-response strength comparable to $u$ and $l$, yet no such line is seen where it could be resolved. Its absence cannot distinguish the models because detection requires optical spin contrast (Sec. S5). Indeed, $q$ itself is 17 times stronger than the median linear-response ratio of the $S=1$ model for a microwave field $b_1 \perp B$.

The fits distinguish the models tested here. The combination rule is consistent with $u$, $l$ and $q$ sharing three levels, which could belong to a larger manifold. None of the tested second-order $S=3/2$ fits hosting $q$ reproduces the positions. A ladder on $u$, $c$ and $l$ with an independent line at $u+l$ remains possible, however. Parsimony and the coincidence premise disfavor it; the strong-line positions do so only if the window shift of $l$ is a constant offset (Sec. S3). We did not test field-linear terms cubic in $S$, which are allowed for $S=3/2$, or models with $S=2$, which is non-Kramers like $S=1$, or $S\geq 5/2$. A sweep through zero field (Sec. S5) is needed to rule out $S=3/2$ without relying on a fit. Full comparisons and limitations are given in Secs. R11 and R20.

## S4. Amplitudes, central line and optical cycle

### Why is $q$ visible?

The measured heights describe cw contrast under microwave power and saturation conditions that were not controlled. Over 331–705 G, the ratio is $A_q/A_u=0.76$, with the ratio falling only from 1.06 to 0.60. Linear response predicts contrast $\propto |\langle i|b_1\cdot S|j\rangle|^2$ for a microwave field along $b_1$. This gives ratios from 0.06 ($b_1 \perp B$) to 0.11 (isotropic $b_1$) for the $S=1$ model, and a decrease by a factor of 4–15 across the field range for both the $S=1$ and $S=3/2[u,l,q]$ models. Linear response therefore fails for both. A one-parameter saturation law fits them equally well. Saturation is plausible: the Rabi frequencies of defect 6 [2], 24.1–39.3 MHz, far exceed the homogeneous widths of 2.8–7.8 MHz inferred from their decay. The optical-cycle models in Sec. S4 replace this static saturation law (Secs. R7, R11 and R8).

### The central line

The central line $c$ is 2.9 times wider than $u$, based on median FWHM over 331–705 G. Within the line-width limit, hyperfine coupling broadens $c$ and $u$ of a shared $S=3/2$ manifold similarly, giving a width ratio of at most 1.1. To produce a line at $q$ in every column, such an $S=3/2$ model would require a nuclear coupling with norm at least 116 MHz by the Weyl bound. For one isotropically coupled nucleus, this would make every $\Delta m_s=1$ line wider than $u$.

The pulsed ODMR spectrum of defect 2 [2] was recorded at a field not specified in the data files. The inferred field is $994\pm4$ G from $c$, provided the map's field scale and direction apply. The spectrum contains $c$, $u$ and $l$, with $u$ and $l$ displaced from their cw tracks by $-7\pm5$ and $-39\pm14$ MHz, respectively (line-fit errors only).

The predicted $q$ lies outside the window. As in the overlapping cw windows (Sec. S2), $l$ shifts while $u$ does not. The pulsed $l$ is marginal, however: its width reaches the fit bound, leaving its position and height dependent on the line model. We infer that the only microwave-prepared transient for defect 2 drives $c$. Luo *et al.* report driving the resonance with the largest contrast, and $c$ has the largest contrast in every defect-2 spectrum. The files provide no frequency, field, power or cycle for that drive.

For defect 6, the Rabi-frequency ratio $c/u = 0.82 \pm 0.01$ ranges from $0.80$ to $0.90$ across envelope models. It lies between the equal-$B_1$ predictions of the axial $S = 3/2$ model ($1.15$) and the $S = 1$ + $c$ model ($0.71$). The lines were driven at different frequencies, though, and Luo *et al.* attribute the differences to the antenna response. Without knowing $B_1(f)$, this ratio cannot select either model. If it were a two-photon resonance between the outer $S = 1$ levels, $c$ would nutate at $0.59$ MHz rather than the measured $24.1$ MHz. None of these central-line results decides among the four main-text interpretations of $c$ (Secs. R14, R8 and R20).

### Optical-cycle models

We used Lindblad master equations for the spin sublevels of the ground, excited and shelving states, including optical pumping, spin-selective intersystem crossing and an explicit microwave field. The models were compared post hoc with the cw map, the $c$-driven fluorescence transients and the photon correlations (Sec. R21). None of the physically consistent fits tested reproduces all three observables with shared intrinsic rates and a common microwave field; this is not an exclusion.

Two model families reproduce both the magnitude of $q/u$ and its weak field dependence. Each requires a condition that the archived data cannot test, and the data cannot separate them. The first is an isolated $S = 1$ under a finite, polarized drive that weakly drives $u$. Its line widths are in tension with the full map, which does not exclude it. We compared its transients and correlations with summaries of the measured data. At the fitted rates, the light-on overshoot, defined as the first bin divided by the plateau, matches the measured value only if the transients were recorded at a small fraction of the cw pump power. Its mean microwave-off bunching over the measured shoulder stays below the background-corrected bound only at a still smaller fraction. These are necessary conditions, rather than fits to those observables. The second family is model B, an $S = 1$ read out through a coupled spin-1/2. It fails to reproduce the prompt rise of the $c$-driven transient unless the microwave-off trace was recorded at lower laser power.

*Power.* Selected fits from the two families predict opposite changes in $q/u$ with microwave power at one field (Table S5). Increasing the power preferentially strengthens the weakly driven $u$ of the isolated-$S = 1$ fits, so $q/u$ falls. In model B it instead strengthens the weakly driven $q$, so $q/u$ rises. These signs characterize the retained fits; they are not bounds on either model family. The searches did not seek fits with the opposite sign, and the drive strength at the defect, its polarization and the antenna response were inferred from the fits (Sec. R21).

**TABLE S5.** Predicted amplitude ratio $q/u$ near $505$ G as a function of microwave power relative to the cw map. For model B, the tabulated predictions use an incoherent line drive. Calculations with an explicit microwave drive support this approximation for $u$, $l$ and $q$, without recomputing the tabulated values. Entries are medians over posterior draws, with 0 dB set separately for each draw at the power that reproduces the measured $q/u$. The narrow 95% interval at $+6$ dB, $1.65$–$1.75$, is the spread across these draws. For the isolated $S = 1$, entries give the best fits for four combinations of model class and drive response, with 0 dB at the fitted drive.

| Model | $-6$ dB | 0 dB | $+6$ dB | $+12$ dB |
|---|---|---|---|---|
| B ($S = 1$ with spin-1/2 read-out) | 0.36 | 0.78 | 1.70 | 2.95 |

| Model | $-6$ dB | 0 dB | $+6$ dB | $+12$ dB |
|---|---|---|---|---|
| Isolated $S=1$, charge-transfer pair | 1.09 | 0.81 | 0.55 | 0.44 |
| Isolated $S=1$, metastable triplet | 1.15 | 0.89 | 0.53 | 0.31 |
| Isolated $S=1$, own excited-state tensor | 1.51 | 0.81 | 0.48 | 0.38 |
| Isolated $S=1$, charge-transfer pair, field-independent drive | 1.06 | 0.80 | 0.56 | 0.38 |

### Low-field positions and angle scans

When fitted at $B \geq 331$ G, the ladder in Sec. S3 predicts that $c$ approaches its Kramers gap (Sec. S5) as $B \to 0$, whereas the measured $c$ continues to decrease. We tested this difference assuming a fixed field direction with a constant stray field of at most 150 G per axis, and continuous tracking of each line. Adding the 15 positions of $c$ and $u$ at 92–311 G raises the total $\chi^2$ of the best ladder found by 95.6 relative to its high-field refit (nominal $p = 8.8 \times 10^{-14}$ for 15 degrees of freedom). The corresponding increase is 26.4 for the $S=1$ + $c$ model. With an additional 20 MHz systematic uncertainty, the increases become 34.5 and 9.8, respectively. Further random starts found slightly lower ladder minima and raised its nominal $p$ value, but every ladder fit still showed a larger increase than the $S=1$ + $c$ model.

This comparison has three limitations. First, 122 of the 129 $\chi^2$ units separating the models when line identities are free come from just four points: the three lowest-field $c$ points and the lowest-field $u$ point. These also move most between extractions. Removing the three lowest $c$ columns leaves the models fitting equally well (40.4 versus 43.5). Second, the ladder becomes viable if the field direction drifts more strongly than a constant stray field permits, by about $50^\circ$ at 92 G relative to 1000 G. In that case the $\chi^2$ increases are 31.5 for the ladder and 51.2 for the competing model, although neither search converged. We know of no mechanism for such a drift, and both window files record continuous field stepping. Third, the minima are not known to be global, the $p$ values have not been calibrated by simulations of the branch-optimized fit, and the high-field fits already exceed their nominal $\chi^2$. The low-field positions therefore disfavor the ladder only under the stated assumptions. Zero-field ODMR with detectable contrast (Sec. S5), or a new measurement with a calibrated three-axis field, would test this conclusion.

The defect-2 angle scans do not yield an acceptable global fit for any tested model. Because the files do not specify the field, we use $c$ to estimate $|B|$. The resulting fits give $\chi^2/dof = 10$ for $S=1$ and 13 for a rhombic $S=3/2$ with free $g$ anisotropy. These scans are inconsistent with an axial $S=3/2$ model, but cannot distinguish $S=1$ from a rhombic $S=3/2$ model. Given the poor fits, we use this comparison only qualitatively. Defect 4 supplies no additional evidence (Sec. S2; Secs. R21 and R9).

## S5. Zero-field predictions and proposed tests

### Why zero field discriminates

We diagonalize Eq. (2) exactly for $S=1$ and $S=3/2$.

For $S=1$, the zero-field lines are $D'-|E|$, $2|E|$ and $D'+|E|$ in the fit frame (Sec. S1). Written without reference to a frame, they are $P_1-P_2$, $P_2-P_3$ and $P_1-P_3$, and therefore satisfy $f_1+f_2=f_3$. The joint fit predicts $0.509 \pm 0.014$, $0.583 \pm 0.017$ and $1.092 \pm 0.005$ GHz, with statistical errors only. Across all fits, the corresponding ranges are 0.50–0.54, 0.56–0.60 and 1.09–1.11 GHz. Neither set includes the window shift of $l$ (Sec. S2), which changes $D'$ and $|E|$. On a calibrated field scale ($g=2$), the line shifts are even in $\pm B$ and scale as $(gB)^2$. The largest joint-fit shifts over all directions are 3.1 MHz at 1 mT and 27.0 MHz at 3 mT. The two lower lines are separated by $|D'-3|E||=$18–86 MHz and may merge, depending on the zero-field width, which the data do not determine. The line at $D'+|E|$ remains resolved.

*The $S=3/2$ prediction follows from Kramers degeneracy.* An $S=3/2$ manifold has one zero-field line, at $2\sqrt{D^2+3E^2}$ between the two doublets. In the fitted $S=3/2[u,c,l]$ ladder at the archived field direction, $u$ and $l$ approach zero frequency, while $c$ approaches this line at 0.97–1.00 GHz. Its components split linearly with field. In this interpretation, $q$ is unrelated and need not approach a Kramers spectrum. At 1 mT, the allowed lines of the axial $S=3/2$ model with $g=2$ span 56 MHz along a principal axis, a lower bound for that model alone. The corresponding spread for a best-fit rhombic $S=3/2[u,c,l]$ model is 97 MHz. Both spreads grow linearly with $B$ and use the field scale of their respective fits; they are slightly larger on a calibrated scale. The zero-field line of the $S=3/2[u,l,q]$ model (Sec. S3) coincides with none of the $S=1$ frequencies.

*A coupled partner changes this prediction.* If $c$ belongs to an $S=1/2$ coupled to the $S=1$, the total spin is half-integer and every zero-field level forms a Kramers doublet. For dipolar couplings $J$ of 2–15 MHz, the largest residual linear slope across coupling orientations and field axes is about 0.22 MHz/mT per MHz of $J$. The sweeps do not bound $J$ because splittings below the line width are unresolved. Distinguishing this case therefore depends on the size of the linear slope.

*These frequency predictions do not guarantee observable contrast.* The spin Hamiltonian does not determine the populations or emission contrast needed to detect a line. Zero-field contrast vanishes for both the $S=1$ and $S=1/2$ lines of an hBN emitter [4], and Luo *et al.* observed no low-field cw contrast in their group-I defect [2]. Failure to detect contrast at zero field therefore falsifies neither model on its own; the branches must be observed and tracked (Secs. R10 and R11, including the zero-field spectrum of $S=3/2[u,l,q]$ and the survey of random $S=3/2$ models).

### Clock transitions and their limits

For the zero-field eigenstates $T_i$ of the isolated non-Kramers manifold, $\langle T_i|S_j|T_i\rangle=0$ for every $i$ and $j$. The zero-field lines are therefore insensitive to $B$ to first order in every direction, with second-order coefficients of magnitude 0.7–3.1 MHz/mT$^2$. This statement applies to the electron spin alone. A hyperfine field shifts $B$ by an amount that depends on the nuclear state and consequently moves the clock point [5]. For quasistatic Gaussian field noise with $\sigma_B=0.1$ mT per component, the model predicts dephasing 104–117 times slower than for a line with linear slope $\gamma$. This factor scales as $1/\sigma_B$. It is an order-of-magnitude estimate for quasistatic magnetic noise only, conditional on an uncoupled partner, rather than a prediction of $T_2^*$ or $T_2$.

Strain and electric fields still act to first order: each line responds to a difference between principal values, leaving these clock transitions vulnerable to charge noise [5]. Electric-field noise from illumination-induced charge fluctuations limits coherence at the NV center's zero-field anticrossing [6], and an optically read GaN center may behave similarly. Because the spin-strain and spin-electric couplings have not been measured, we make no quantitative sensing claim. The estimates above cannot be interpreted as device or coherence gains without measurements of $T_2^*$, the hyperfine couplings and the noise spectrum (Sec. R10).

**Proposed measurements, in order of priority**

(i) *Zero field.* Sweep defect 2 continuously from $-3$ to $+3$ mT and back, preserving one field history and covering all predicted frequencies at low microwave power. Use a filtered drive and check it on a spectrum analyzer for harmonics and intermodulation. Reversing $B$ determines only the residual field along the sweep axis. Transverse fields, including the Earth's field, must be nulled separately, for example with three-axis coils. We specify the matching criteria in advance: three lines, or two if the lower pair is unresolved; each frequency within two statistical errors of the joint-fit value or within the range across all fits, whichever is wider; $f_1+f_2=f_3$ within the measured line widths; and continuous connection of each line to $u$, $l$ or $q$. The shifts must be even in $\pm B$ and no larger than those given above; these criteria do not include the window shift of $l$. If $u$ and $l$ instead approach zero frequency or split linearly when tracked continuously from high field, the isolated $S=1$ is falsified. A small linear term would point to a coupled partner (Sec. S5).

*(ii) One window.* Measure $u$, $l$ and $q$ in one frequency window, or in interleaved windows, to remove the anomaly in $l$ (Sec. S2). If $l$ were free of a window shift, the present positions would already reject the ladder with an independent line ($p \leq 0.005$, none of $200$ replicates).

(iii) *A second emitter.* Repeat test (i) on a group-II emitter with $u+l \neq 2c$, unlike defect 3 (Sec. S2).

(iv) *Two-tone ODMR.* Pump $u$ while probing $l$ or $q$. Their contrast should change if the transitions share a level. A line that shares no level with $u$, such as $c$ in the $S=1$ + $c$ model or $q$ in the ladder, remains unchanged only if its spin and optical cycle are dynamically independent of those of $u$. If $c$ is a coupled spin-1/2 (model B) or a spin pair reached through charge transfer, pumping $u$ can alter its populations and contrast without a resolved splitting. We have not computed this response. Such a change would demonstrate dynamical coupling without establishing a shared level. In the ladder, pumping $u$ acts on $c$ while leaving a dynamically independent $q$ unchanged. The test shows shared levels; it does not by itself show $S=1$.

(v) *Power dependence.* Measure the contrast, width and Rabi frequency of all four lines from below to above saturation, beginning with $q$ and $u$. Below saturation, one-photon contrast grows linearly with power and two-photon contrast grows quadratically. This tests whether $c$ or $q$ has a multiphoton origin. Near $505$ G, at 6 dB above the cw-map power, $q/u$ falls to at most $0.56$ in the best isolated-$S=1$ fits, but rises to $1.70$ in model B, with a range of $1.65$–$1.75$ (Table S5). This comparison tests the selected fits, rather than the model families as a whole (Sec. S4).

*(vi) Equal-frequency Rabi measurements.* Tune $B$ separately for each line so that $l$, $c$, $u$ and, ideally, $q$ can all be driven at the same microwave frequency. The frequency dependence of $B_1(f)$ then drops out of the comparison. At equal $B_1$ and averaged over drive polarization, $q$ in the $S=1$ model nutates $3.0$–$4.1$ times more slowly than $u$. The isolated-$S=1$ fits instead require a polarized drive that drives $u$ more slowly than $q$. The measured ratio would distinguish these interpretations.

(vii) *Optical channel.* Record the zero-phonon line and photoluminescence spectrum while driving each resonance. This would help distinguish a second emitter from another spin or charge configuration of the same emitter.

(viii) *Low-field map.* Record a dense map at low microwave power around the avoided crossing $B^*$ (Sec. S3). This would test branch continuity, though it is less fundamental than test (i).

Full descriptions of the proposed tests and the double-quantum line are given in Sec. R10; the power predictions are detailed in Sec. R21.

## S6. Electronic-structure calculations

The electronic-structure calculations discussed in Sec. VI of the main text disfavor the states examined, but do not identify the emitter.

### Static hole pairs

*The measured tensor constrains the center's symmetry.* A center with $C_{3v}$ symmetry about $c$ would have $E=0$ and its principal axis along $c$. The large measured $E$ rules out this symmetry and instead points to $C_{1h}$ or lower symmetry. At a single field direction, the orientation is constrained only through $n \cdot D \cdot n$ (Sec. S3). We therefore compare computed orientations at the approximate field direction reported by Luo *et al.*, using the tilts from $c$ given in Sec. R16 and the best of the six $a$-type azimuths. We use two tests. The first is the ratio $\rho = n \cdot D \cdot n / D$, with the signed conventional $D = 3/2\, P_k$ defined from the principal value $P_k$ of largest magnitude. This ratio is unchanged by reversing the overall tensor sign. The measured value is $\rho = -0.59$, although its sign depends almost entirely on labeling because $|P_1|$ and $|P_3|$ differ by only $4\,\%$. The second test avoids this labeling issue: it compares the angle between $B$ and the nearer of the two large principal axes, which is $11\text{–}17^{\circ}$ for the measured tensor.

A Ga vacancy leaves spin-$1/2$ holes in the $2p$ orbitals of its neighboring N atoms: one axial N along $c$ (A) and three basal N atoms (B0, B1 and B2). In a complex, partners X (O, C or H) sit on or next to these sites. The tensors, calculated with the Heyd–Scuseria–Ernzerhof (HSE) hybrid functional at exact-exchange fraction $\alpha = 0.31$ in a $92$-site cell (Sec. S7), include only the spin–spin interaction.

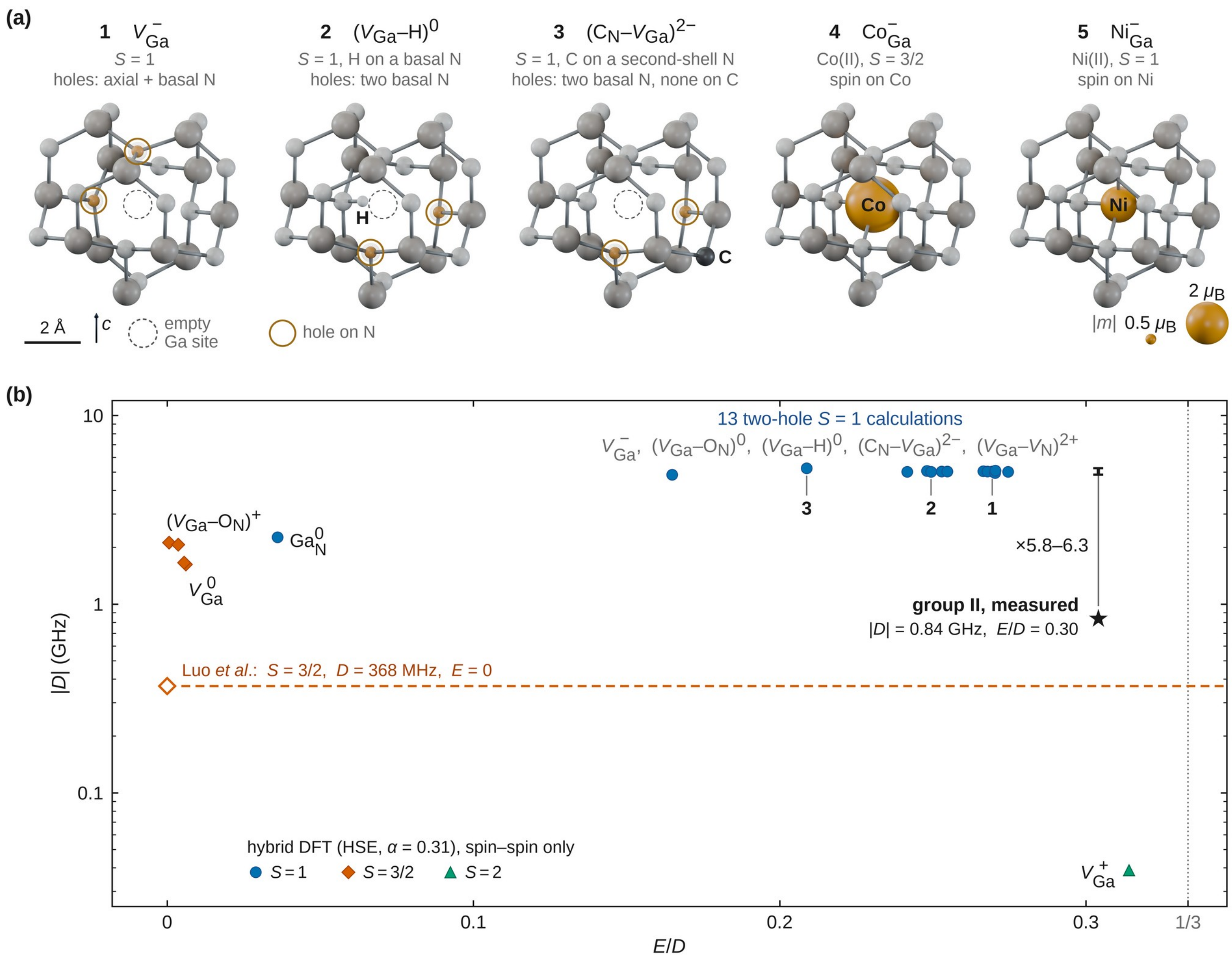

**FIG. S3.** None of the computed tensors matches the measured tensor. (a) Relaxed configurations with $c$ pointing upward. Large and small spheres denote Ga and N, respectively; H, C, Co and Ni are labeled. Gold spheres mark every site with $|m| \geq 0.25\,\mu_B$, with radius proportional to the site moment. Gold rings of a fixed size identify the N atoms carrying a hole. Structures are 1, $V_{Ga}^-$; 2, $(V_{Ga}–H)^0$; 3, $(C_N–V_{Ga})^{2-}$, initialized with C in the second shell and relaxed to the pair in structure 2; 4, $Co_{Ga}^-$; and 5, $Ni_{Ga}^-$. (b) $|D|$ plotted against $E/D$ for the 19 calculations with a spin–spin tensor. The star marks the measured tensor; the diamond and dashed line show the axial $S = 3/2$ model of Luo *et al.* The broken-symmetry basal divacancy and the metal ions are not shown (Sec. S6).

We completed 13 calculations of localized two-hole $S=1$ states, representing 12 distinct states of $V_{Ga}^-$, $(V_{Ga}\text{-}O_N)^0$, $(V_{Ga}\text{-}H)^0$, $(C_N\text{-}V_{Ga})^{2-}$ and the axial divacancy. Their splittings span $|D| = 4.84$–$5.24$ GHz, or $5.8$–$6.3$ times the measured value. The N–N pairs have $E/D = 0.21$–$0.27$, with the principal axis along the hole pair (Fig. S3). Including tensors from a $192$-site cell and the fully hard PAW set gives a range of $4.77$–$5.24$ GHz. Both orientation tests fail for axial–basal pairs (ax+bas) and basal–basal pairs (bas+bas) among the first-round states. These static states are therefore disfavored at this level of theory. Other charge states, delocalized holes, larger cells beyond the one family tested, and spin–orbit coupling remain open possibilities.

The origin of $|D|$ matters for this comparison. The classical dipolar coupling between the two computed hole spin densities is $1.4$–$1.8$ GHz, with $E/D \leq 0.05$. This contribution alone makes the splitting of a static nearest-neighbor hole pair too large and its tensor too axial. Point dipoles would need a separation of $4.5$ Å to reproduce the measured $|D|$, greater than the distance between nearest-neighbor holes. The non-classical remainder accounts for about two thirds of $|D|$ and nearly all of $E$, but has not been benchmarked. It is unclear whether it reflects exchange and spin polarization or an artifact of spin contamination. Three further PAW data sets change $|D|$ by at most $4.8\,\%$; using $\alpha = 0.25$ changes it by $8.9$–$10.2\,\%$. These checks rule out a data-set artifact only for the sets examined. The conclusion for static pairs holds regardless of the origin of the non-classical contribution, while the averaging results below depend on it.

No other computed state matches both the spin and the tensor shape. The spin–spin tensor of $Ga_N^0$ ($S = 1$, with spin on Ga) does not match the measurement. Spin–orbit coupling is especially important for a spin on Ga, however, and was not computed, so this state is not excluded (Secs. R16 and R17).

**Averaging by hole hopping**

Fast hopping replaces the large tensors of individual hole patterns with a weighted mean constrained by symmetry. We model the classical contribution using pair tensors, with point dipoles placed along the N–N vectors connecting the holes.

(i) Patterns related by $C_3$ have equal populations at equilibrium unless a $C_3$-breaking perturbation, such as strain or a neighboring defect, splits their energies. Such perturbations are not included here. Any set of hops closed under $C_3$ then gives $E = 0$ by symmetry in either tensor model, regardless of the relative populations of different $C_3$ orbits. Hopping within $V_{Ga}^-$, or within a complex with X on the axial N, therefore cannot reproduce the measured rhombicity under these conditions. Unequal populations within one $C_3$ orbit leave its $cc$ component unchanged, keeping $|D|$ at or above $3.17$–$3.38$ GHz for the computed tensors. A $V_{Ga}^-$ set supplemented with bas+bas patterns can meet the comparison tolerances when their tensors are borrowed from the X-axial complexes. The pair average over bas+bas patterns with X axial can also meet them, but its $P_1$ axis remains along $c$. It passes the label-free test only when $B$ lies about $11$–$17^{\circ}$ from $c$, slightly beyond the approximate $10^{\circ}$ reported by Luo *et al.* An HSE calculation started without the PBE step converged to a $V_{Ga}^-$ state with two basal holes. This stationary solution is excluded from the counts and orbits above and does not establish metastability or hopping kinetics. Its tensor lies within the static ranges; its energy order relative to one matched ax+bas pattern is not established. With this tensor, refitted populations pass every test of this section, but only with an uncomputed bias between basal sites and a mean-energy difference between the ax+bas and bas+bas patterns, which the single static energy difference neither supplies nor excludes (Sec. R16).

(ii) With O or H on a basal N, hopping between the mirror partners A+B1 and A+B2, degenerate within $4.1$ meV in this cell, reproduces $E/D$ and $\rho$. However, $|D|$ is $4.3$ times too large, the agreement depends on the unbenchmarked non-classical contribution, and pair tensors predict the wrong orientation. (iii) B1+B2 is the lowest pattern in this cell, lying $18.3$ meV below A+B1 for the H complex and $27.1$ meV below A+B2 for the O complex. This ordering also holds at $\alpha = 0.25$ and, for $(V_{Ga}\text{-}H)^0$, in the $192$-site cell. At $295$ K, the three-pattern average consequently loses the measured shape. No population lowers $|D|$ below $3.44$ GHz ($3.39$ GHz in the larger cell), and retaining the shape requires an ordering error of at least $57$–$68$ meV. (iv) With pair tensors, satisfying all three targets requires a shift of $46$–$61$ meV. (v) Nudged-elastic-band estimates in this cell, without a climbing image, give barriers near $0.11$ eV for hopping into B1+B2 and $0.12$ eV for the mirror hop. At an assumed attempt frequency of $10^{13}$ s$^{-1}$, these barriers are too high to produce narrow lines with the computed tensors, although their $|D|$ already rules out averaging. In a test of kinetics alone, with tensors reduced by a factor of $4.3$ and B1+B2 unpopulated, the mirror hop produces narrow lines above about $3$–$5 \times 10^{12}$ s$^{-1}$.

The kinetics neither establish nor exclude motional averaging; the screen gives it no support, and the computed lowest-energy pattern works against it (Secs. R16 and R17).

**Limits of the screen**

(i) The cell lacks $C_3$ and mirror symmetry. A $192$-site cell changes the $\left(V_{Ga}\text{-}H\right)^0$ tensors by $-4.9\,\%$ and $+3.6\,\%$, but a comparison of two cells does not establish convergence, and no $k$-point check was performed. (ii) The calculations use only the $\Gamma$ point, with no finite-size correction to the pattern energies. Errors of $46$–$68$ meV would change the conclusions about averaging. (iii) The $\alpha=0.25$ runs were initialized from the $\alpha=0.31$ solutions. They test the ordering and tensors of those seeded patterns, but do not establish whether localization persists independently of the starting state. (iv) The PAW comparisons measure sensitivity to the data set; they do not bound the error. Hard O and C sets were not tested. (v) Spin purity remains unestablished. For $\left(V_{Ga}\text{-}H\right)^0$ on A+B1, an unrelaxed broken-symmetry $M_S=0$ determinant evaluated at the triplet geometry lies $+9.8$ meV above the triplet in this cell. No closed-shell singlet or other pair was tested. (vi) Spin–orbit coupling was not computed for the $2\,p$-hole pairs and may be comparable to $E$. The shape and orientation tests include only spin–spin coupling. (vii) Charge states were tested in part. We evaluated charge-transition levels of $V_{Ga}$, $V_{Ga}\text{-}O_N$ and $C_N$, including charged-defect corrections, from the existing hybrid runs in the $92$-site cell at $\alpha=0.31$ and the $\Gamma$ point alone. The comparison covers only the computed high-spin states: spin ordering is not bounded, and the tested sensitivities do not constitute physical error bounds. If the Fe acceptor pins the Fermi level in the upper part of the gap, as reported for Fe-doped semi-insulating GaN [7], the two-hole $S=1$ states $V^-_{Ga}$ and $\left(V_{Ga}\text{-}O_N\right)^0$ are not the stable dark charge states among those computed. For $V_{Ga}$, the published levels give the same conclusion [8]. In such a sample, these states would have to be photogenerated, occur where the local Fermi level is lower, or differ from the states computed here. Adiabatic threshold estimates permit sequential photocharging into them under $532$-nm excitation, but also into $V^0_{Ga}$ ($S=3/2$). We did not compute vertical thresholds, cross-sections, rates or populations. Our (−/2−) and (2−/3−) levels of $V_{Ga}$ lie closer to the valence-band maximum than the published levels by more than our preset tolerance; the reason remains unresolved. The (+/0) and (0/−) levels pass this comparison. Formation energies and concentrations, convergence in cell size beyond the correction, in sampling and in the functional, the other complexes and the transition-metal charge states of Sec. S6 remain untested. (viii) Not computed: the dependence of the hop barriers on $\alpha$ and cell size, $g$ tensors and optical transitions. The hyperfine tensors were computed but not analyzed. These calculations therefore give no conclusion about $g$ or the hyperfine couplings. The charge levels constrain only adiabatic charging energies, not optical spectra or rates.

Several classes remain unscreened. For the four Fe complexes $\left(Fe_{Ga}\text{-}O_N\right)^0$, $\left(Fe_{Ga}\text{-}V_N\right)^0$, $\left(Fe_{Ga}\text{-}H_i\right)^0$ and $\left(Fe_{Ga}\text{-}V_{Ga}\right)^{2-}$, the lowest computed spin sector is $M_S=2$. The $M_S=1$ sector, which would be lowest for an $S=1$ ground state, lies $1.0$–$1.4$ eV higher in all four cases at the level of the final run. These calculations compare spin sectors only. We did not extract $D$ or test $S=2$ against the spectrum. They do not screen Fe-related centers as a class, nor N vacancies, interstitials or mixed H, O and C complexes (Secs. R16 and R17).

**Transition-metal ions**

For a $3d$ ion, the zero-field splitting arises mainly from second-order spin–orbit coupling, which is absent from the screen above. We considered four substitutional ions (Table S6) using two approaches. Route A, a periodic estimate, gives only $0.30$ times the measured $Cr^{4+}$ splitting and fails its calibration. Its values therefore cannot be treated as predictions. Route B uses multiplet calculations in an embedded cluster (Sec. S7). The small-cluster Cr4+ result, including spin–spin coupling (SSC), meets the preset $\pm 30\,\%$ criterion, and it lies near the upper limit. The large-cluster estimate, with the spin–spin contribution transferred from the small cluster, is $1.43$ times the measured splitting and falls outside this criterion (Sec. R18). $Cr^{4+}$ is the sole calibration anchor, its assignment is disputed, and the Fe3+ test fails. The method's accuracy therefore remains unestablished, and its calculated $E$ and tilts are noise.

**TABLE S6.** Route B results for substitutional ions, with one charge and spin state considered for each ion.

| Ion and state | Role | Route-B outcome | Verdict |
|---|---|---|---|
| $Cr_{Ga}^{+}$ ($Cr^{4+}$, $S=1$) | calibration | small cluster with SSC: passes; $1.297$ times measured $\lvert D\rvert$ | excluded (spectrum of assigned center) |
| $Fe_{Ga}^{0}$ ($Fe^{3+}$, $S=5/2$) | calibration | fails, $0.096$ of measured $D$ | not this center (measured $D$, $E=0$) |
| $Co_{Ga}^{-}$ ($Co^{2+}$, $S=3/2$) | candidate | smallest gap $11.9$ GHz, not size-converged | disfavored, not excluded |
| $Ni_{Ga}^{-}$ ($Ni^{2+}$, $S=1$) | candidate | no $S=1$ spin Hamiltonian | disfavored, not excluded |

*$Co_{Ga}^{-}$*. A transition within a Kramers doublet of an $S=3/2$ manifold starts at zero frequency, whereas the inter-doublet gap at zero field is $\Delta_0=2\sqrt{D^2+3E^2}$ (Sec. S5). High-field extrapolated intercepts do not determine which zero-field branches the observed lines approach. The best $S=3/2$ fits with fixed assignments have $\Delta_0\leq 2.41$ GHz (Sec. S3), but these are fitted values from rejected models, not a bound. Weyl's inequality provides a necessary upper limit of $\Delta_0\leq 8.8$ GHz for $\lvert g^T\hat{n}\rvert\leq 2.4$, assuming one center, three distinct single-photon electron transitions, and a field dependence containing only the linear Zeeman term on the uncalibrated file field scale. Measured $d^7$ analogues suggest $\lvert D\rvert$ values of tens of GHz: $Co^{2+}$ in ZnO has $D=+82.7$ GHz [9], and $Ni^{3+}$ in GaN provides another analogue [10]. We found no measured $D$ for $Co^{2+}$ in GaN, so this comparison remains an inference. Route B gives gaps of at least $11.9$ GHz, but they decrease in the larger of the two clusters, have not converged with cluster size and omit spin–spin coupling. They are therefore not lower bounds. These comparisons disfavor this candidate without excluding it. The case rests mainly on the measured analogues; the unconverged route-B results provide supporting evidence rather than a definitive test.

*$Ni_{Ga}^{-}$*. At the static hybrid-functional geometry, the splitting of the orbital triplet by $C_3$ breaking is much smaller than the spin–orbit splitting of $^3T_1$. Route B gives a non-magnetic spin–orbit ground level and projection norms of $0.09$–$0.52$, well below $0.9$, so an $S=1$ spin Hamiltonian does not apply. In this static description, $Ni_{Ga}^{-}$ does not reproduce group II. It is disfavored rather than excluded because we did not compute a vibronic (Jahn–Teller) treatment, a $C_{3v}$-constrained geometry, a large cluster, spin–spin coupling or another charge state.

Within the calculations performed under the $S=1$ interpretation, none of these four ions accounts for the center. This comparison does not screen transition metals as a class. The ion-specific results are given in Sec. R18, and the spin–orbit methods and their known biases in Sec. R19.

## S7. Methods, cross-checks and data

Sections S1, S3, S4 and S5 describe the line extraction, fitting procedures, optical-cycle models and spin-Hamiltonian calculations. For the electronic-structure screen, we used VASP 6.6.1 [11,12] with the Heyd–Scuseria–Ernzerhof hybrid functional [13–15], exact-exchange fraction $\alpha=0.31$ [8] and screening parameter $0.2$ Å$^{-1}$. Projector augmented waves [16,17] were used with a $400$ eV cutoff in a $92$-site supercell at the experimental lattice, sampled at the $\Gamma$ point alone. We evaluated the spin–spin tensor from the collinear spin density using `LDMATRIX` [18]. For transition metals, route A adds a spin–orbit contribution through the force theorem [19], using noncollinear runs on a frozen PBE+$U$ density with $U=3$ eV on $3d$ [20]. Route B uses

ORCA 6.1.1 [21] for an embedded cluster at the hybrid-functional geometry, with state-averaged CASSCF in the ($n_d$,5) active space of the $3d$ shell, NEVPT2 [22], and quasi-degenerate perturbation theory with the mean-field spin–orbit operator [23]. Sections R2, R16, R17, R18, R19 and R21 give the full settings, larger-cell and PAW tests, and comparisons between code builds.

The primary line-extraction and spin-Hamiltonian fitting routines were cross-checked with alternative scripts to assess numerical consistency. Electronic-structure post-processing was checked by recomputing the reported quantities from the original output files (Sec. R22).

### Data and code availability

The raw data are from Ref. [1] (doi:10.7298/azf0-hc03, CC BY-NC 4.0). The line tables, the outputs of the electronic-structure calculations and a table giving the source of every printed number are archived at Zenodo (doi:10.5281/zenodo.23744355). The other analysis materials are available from the author on request. They include the analysis code, software environment and random seeds, the significance search and its null maps, the cross-check scripts with their outputs, the electronic-structure inputs, the zero-field, clock-transition and double-quantum calculations, figure data, and scripts that regenerate the figures and, from the stored results, every printed number (Sec. R23).

## References for Part A

## Part B: Detailed analyses (Secs. R1–R23)

### R1. Data and field scale

The sample is a $4\,\mu$m Fe-doped semi-insulating HVPE GaN layer on sapphire, excited at $532$ nm with about $20\,\mu$W [B1]. The defect-2 central line has an intercept of $-4\pm4$ MHz. In defects 3 and 4 it has $g_{file}=1.914\pm0.005$ and $2.009\pm0.004$, with intercepts of $+58\pm7$ and $-4\pm6$ MHz.

### R2. Line extraction

The systematic variants are Lorentzian or Voigt satellites, a Lorentzian or Gaussian $c$, no replica, stitched windows, a quadratic baseline and local windows. Tracks follow the offsets $u-c$, $c-l$ and $q-2c$. These offsets are tracking coordinates rather than frequency identities; in particular, the offset of $q$ from $2c$ varies with field. Joint-fit positions are bounded around the track seeds. In defect 2, $l$ crosses $c/2$ just below the combination columns.

### R3. Same-column combinations

*Separation cut.* In order, the columns of Table S1 have $|l-c/2|=4$, 94, 118, 163, 167, 204, 217, 232 and 267 MHz. Cuts at $80$, $100$ and 120 MHz keep $8$, $7$ and 6 columns and give $q-(u+l)=4.4\pm5.9$, $8.6\pm4.8$ and $10.4\pm5.3$ MHz; they differ because the $560$ G column contributes $-25$ MHz. The fitted estimates include $9.1\pm4.1$ and $6.3\pm6.2$ MHz at the $80$ MHz cut. At $100$ MHz the corresponding values are $10.1\pm4.6$ and $9.9\pm5.8$ MHz, statistical errors only; with the line-shape and window terms, each $100$ MHz value is about one standard deviation from zero. Omitting the replica and retaining blended columns gives $13.9\pm3.9$ MHz; $q-(u+l)$ changes sign where $l$ crosses $c/2$, the signature of a blend.

The difference $q-2c$ falls from $165$ MHz at $331$ G (with a maximum of $168$ MHz at $351$ G) to $58$ MHz at its minimum. The replica of $q$ at $q/2$, inside the width of $c$, is detected ($\Delta\chi^2=39$); modeling it moves $c$ by up to $7$ MHz in one column, $q-2c$ by $+2$ MHz and $q-(u+l)$ by $-0.8$ MHz.

The uniform-offset test for the lower-window $l$ positions, including its limitations, is described in Sec. R12.

*Defect 3.* The error of $q-2c$ is propagated from per-column errors of $19$–$53$ MHz; the scatter, $\pm3$ MHz, understates it. Re-extracting the line positions gives $q-(u+l)=11\pm6$ and $q-2c=-13\pm17$ MHz on $5$ columns at $600$–$720$ G.

### R4. Spin-Hamiltonian fits

*Orientation.* $B$ lies $79$–$90^{\circ}$ from the conventional $z$ axis. We take the field direction of Luo *et al.* (about $10^{\circ}$ from $c$, basal projection along $a$) as exact [B1]. A $C_{1h}$ center has one principal axis along its mirror normal, an $a$ direction at least $90^{\circ}-10^{\circ}$ from $B$. This axis therefore cannot be $P_1$. In the three mirror planes, $n\cdot D\cdot n=0.491$ GHz gives $6$ tilts of $P_1$ from $c$: 2, 7, 8, 9, 20 and 24 (2–$24^{\circ}$ over all fits).

For the ladder model, none of $150$ bootstrap refits of $S=3/2[u,c,l]$ misses $q$ by less than $605$ MHz rms. Fitting the strong lines gives $\chi^2=288$ ($S=1+c$: 221; 145 points, $6$ parameters each); point errors omit correlated systematics, so this difference alone cannot distinguish the models.

## R5. Branch significance and curvature

$T$ sums the Gaussian-filtered residual map (80 MHz full width) along straight tracks, divided by $\sqrt{N}$ for $N$ columns and maximized over $g_{eff}=3$–$5$. A linear fit to $q$ gives $g_{eff}=3.74\pm0.02$. The $S=1$ asymptote $2g\mu_B B$ matches the $g=4$ guide line [B1]; neither intercept is used as evidence.

## R6. Prediction of $q$ and the avoided crossing

The predicted gap is centered at $1.500\pm0.005$ GHz; Luo *et al.* reported the crossing near $300$ G and $1.5$ GHz [B1]. Refitting gives $B^*=255\pm3$ G and a gap of $265\pm18$ MHz. The observed minimum is the smallest $q-u$ among the 5 relaxed-detection columns at $228$–$311$ G. At $249$ G the $S=3/2[u,l,q]$ fits predict $q-u=294$ MHz ($g$ free) and $133$ MHz ($g$ in 1.8–2.2).

## R7. Line amplitudes

In linear response, the $S=3/2[u,l,q]$ model predicts $A_q/A_u=0.16$–$0.46$ over its free-$g$ family and $0$–$0.30$ with $g$ in 1.8–2.2. Luo *et al.* used a microwave power of about $20$ dBm [B1]. No $S=3/2$ $\Delta m_s=2$ line appears at $u+c$ or $c+l$ (one-sided 95% bounds: $17\%$ of $A_q$ per position, $6\%$ combined).

## R8. Rabi data of defect 6

Damped-cosine fits give, for $l$, $c$ and $u$, $\Omega/2\pi=39.29\pm0.05$, $24.1\pm0.4$ and $29.23\pm0.06$ MHz and Rabi-decay times $T_{2,R}=108\pm4$, $41\pm5$ and $113\pm6$ ns. These decay times are distinct from $T_2^*$ and $T_2$. The predicted ratios are averaged over azimuth for $b_1\perp B$; the two-photon value comes from an exact Floquet calculation.

## R9. Angle scans

A single laboratory-to-defect rotation is used for all spectra. With $57$ degrees of freedom, $\chi^2=596$ ($S=1$), $763$ (rhombic $S=3/2$) and $2857$ (axial). When $c$ is used as the field gauge, the $S=1$ parameters are degenerate. Defect 2 also switches between spectral variants [B1].

## R10. Zero field, double-quantum line and tests

At $1$ mT the $S=3/2[u,l,q]$ spectrum spreads by $57$ MHz (strong components, free $g$) and $62$ MHz ($g$ in 1.8–2.2). The survey of randomly sampled models is described in Sec. R11. For $\sigma_B=1$ mT the protection factor is $10$–$12$; magnetic high-field widths would imply $\sigma_B=1.0$–$1.5$ mT and a factor of $7$–$10$. At zero field the lines have $(\partial_{D'}f,\partial_E f)=(1,-1)$, $(0,2)$, $(1,1)$; tensor rotations enter at second order.

For the double-quantum line at the measured direction and $70$–$100$ mT, $\partial f_q/\partial D'=0.03$–$0.07$, compared with $0.79$–$0.94$ for $u$ and $l$. For equal, independent $D'$ and $E$ noise, $q$ has 6–9 times less equivalent field noise than the better single-quantum line. This order-of-magnitude estimate applies away from $B_c$. For $B\parallel z$ only the $D'$ channel is removed [B2–B4]; $E$ still enters at first order. There, at $30$–$100$ mT, $B_1\parallel B$ drives $q$ at $0.40$ to $0.14$ times the single-quantum Rabi frequency at equal $B_1$.

*Tests.* (i) The sweep must also cover $2.28$–$2.41$ GHz ($S=3/2[u,l,q]$). (iv) Autler–Townes splitting needs a Rabi frequency above the linewidths. Pumping $q$ should split $u$ and $l$; a pair's $T_+\leftrightarrow T_-$ line would split the lines under $c$. (v) Under the saturation law, $q$ saturates at 9–16 times the power of $u$.

## R11. An $S=3/2$ hosting $u$, $l$ and $q$

With free $g$, the fit Jacobian has rank 4 for both models, compared with nominal parameter counts of 5 and 6. Using these alternative counts changes $\Delta BIC$ by 5. The nearest-transition fit provides a bound on every assignment only if the search has found the minimum of its discontinuous objective. The searches were not certified to have found global minima. These numerical comparisons therefore apply to the fitted solutions and do not constitute an analytic exclusion of the full Hamiltonian family. With a constant lower-window offset of $l$ the fitted offset for $S=3/2$ reaches the $\pm 60$ MHz bound ($S=1$: $-2.3$ MHz).

The fitted amplitude of the fourth-level line relative to $q$ is $0.045\pm0.026$ ($g$ in 1.8–2.2; 95% limit 0.087) and $-0.040\pm0.045$ (free $g$; limit 0.074), compared with 1.0–1.4 if detected in the same way as $u$ and $l$. For free $g$, the line lies within 150 MHz of $q$ in 14 of 21 $q$ columns, where it cannot be tested separately.

At zero field, the best $S=3/2[u,l,q]$ fits each have one line, at $2.28\pm0.04$ GHz ($g$ in 1.8–2.2) or $2.41\pm0.05$ GHz (free $g$). Among 16000 randomly sampled $S=3/2$ models with $g$ in 1.8–2.2, none has a spread below 42.7 MHz at 1 mT when components $\geq 5\%$ of the strongest matrix element are included. These sampled results do not establish a lower bound for the model family.

*Intensities.* With a free scale, linear response fits the 21 ratios $A_q/A_u$ with $\chi^2=131$–193 ($S=1$) and 51–159 ($S=3/2$) for 20 degrees of freedom. The saturation law, contrast $\propto s/(1+s)$ with $s=\kappa|M|^2$, gives $\chi^2=24$–25, except for $g$ in 1.8–2.2 with $b_1\perp B$ ($\chi^2=443$), where $q$ has no matrix element.

## R12. Field offset between the frequency windows

Removing $q-2c$ needs $\delta B\leq-10.3$ G with the measured slope of $c$ ($-10.6$ G with the nominal slope $2g\mu_B/h$). Such a shift is excluded at about $7\sigma$ by the common offset $\delta B=+0.3\pm1.5$ G inferred jointly from the three crossing lines (Sec. S2). The $4.4\sigma$ exclusion quoted in Sec. S2 and the main text uses the central line alone for an offset of $-10$ G.

Where $l$ crosses the overlap, the lower-window $u/2$ replica shifts by $-2.9\pm2.8$ MHz relative to 467–722 G; a field offset predicts $-18$ MHz. Candidate causes are switching of spectral or charge configuration between acquisitions (defect 2 switches between spectral variants [B1]), consistent with the changed amplitude, or pulling of $l$ by the strong drive. In the $S=1$ fit, allowing a uniform offset of the 18 lower-window $l$ positions gives $-2.3\pm2.4$ MHz. Fixing the offset at $-31$ MHz raises $\chi^2$ by 143. This test is partly circular because it assumes the $S=1$ model and uses $q$. It does not constrain shifts that vary between columns.

## R13. Line shapes, replicas and significance

Removing the replicas $c/2$, $u/2$, $l/2$, $q/2$, $c/3$ and $u/3$ one at a time raises $\chi^2$ by 103, 58, 25, 39, 12 and 10, against about $10\pm5$ for an absent replica. The $q/2$ replica lies inside $c$ ($c-q/2=-51$ MHz) and could pull it; modeling it gives $q-2c=104.5\pm7.4$ MHz.

Null-map residuals are resampled in two-dimensional blocks (2000 maps), frequency blocks or a wild bootstrap (1000 each). Each map undergoes the full search of 45 statistics per defect: three residual and two raw map variants, each searched with three kernels and three track families. The Westfall–Young $p$ for defect 2, adjusted over this full family, is $0.010$–$0.024$ across resampling schemes; the extrema in Sec. S2 refer only to the 27 residual-map statistics. These estimates have finite Monte Carlo uncertainty and do not cover every line-shape or cut choice. The extraction finds a false $q$ in 0.5% of null-map columns.

## R14. Hyperfine

Electron–nuclear diagonalization covered 69 columns, four electronic models, up to two Ga or $^{14}$N nuclei and 2000 random tensors per model. Unresolved hyperfine within 100 MHz widths caps $|A|$ at 15–50 MHz; no transition lies within $\pm 40$ MHz of $q$. Weyl's inequality places these transitions within $2\,\|V\|$ of the electronic transitions. $q$'s slope, $5.23 \pm 0.03$ MHz/G, requires $2 S g \geq 3.5$; $S = 1/2$ with $g = 2.1$ gives at most $2.94$ MHz/G. Hyperfine shifts the $S = 1$ $q - (u + l)$ by $-5.1$ to $+1.9$ MHz ($\leq 7.1$ MHz for random tensors).

## R15. Spin pair versus double-quantum line

For pair $\Delta m_s = 1$ lines at $c_1 > c_2$ under $c$, with weights $w_1$ and $w_2$, $q - 2c \simeq (c_1 - c_2)(w_2 - w_1)/(w_1 + w_2)$. A pair needs a splitting above $q - 2c$ (58–168 MHz), with the lower line clearly stronger. In hBN, the refitted $q - 2c$ is $+28 \pm 5$ MHz at $66$ mT and $+79 \pm 28$ MHz at $43$ mT. The $S = 1$ calculation uses $D = 950$ MHz and $E = 0.10$ GHz in our convention (their "$E = 200$ MHz") [B5]. Whitefield *et al.* favor the pair partly because $S = 1$ level-crossing features are absent. In GaN the $q - u$ minimum appears at the predicted field, in one relaxed-detection column.

## R16. Density-functional screen

*Settings.* Standard VASP PAW sets (Ga with $3d$ in the valence) and NUPDOWN $= 2S$ were used. Each calculation ran a PBE [B6] step from seeded holes, a hybrid relaxation, a final hybrid run (spin–spin [B7] and hyperfine tensors) and a released-spin run. The released-spin run left the moments unchanged and changed each energy by at most $0.0002$ meV. Because the released-spin calculation starts from the same wave function, the result establishes only self-consistency; the ground state remains undetermined. Every defect is $C_1$ in the cell, so site operations that should leave a state unchanged change its tensor by 31–132 MHz. Repeats of one state differ by $\leq 2.5$ meV.

*Builds.* Frontier GPU and local CPU builds ran the screen, an Anvil CPU build the transition-metal calculations. At one $V_{Ga}^{0}$ geometry, the Frontier (fresh start) and Anvil (restarted from the local wave function) builds reproduce the local tensor to $0.31$ and $0.05$ MHz, hyperfine couplings to $0.013$ and $0.002$ MHz and energy to $0.75$ and $0.01$ $\mu$eV.

*Orientation.* For $B$ $10^{\circ}$ from $c$ with its basal projection along $a$, the measured $P_1$ axis lies 1–27$^{\circ}$ from $c$ (6–22$^{\circ}$ or 0–32$^{\circ}$ for $B$ $5^{\circ}$ or $15^{\circ}$ from $c$) and the conventional axis 68–90$^{\circ}$, assuming no site symmetry. Computed main axes lie 35–36$^{\circ}$ from $c$ for ax+bas, 89–90$^{\circ}$ for bas+bas and $80^{\circ}$ for C–N pairs; $\rho$ is $+0.30$ to $+0.31$ for ax+bas and $-0.50$ to $-0.44$ for bas+bas pairs, which reach the measured $\rho$ only for $B$ at least 23–33$^{\circ}$ from $c$; C–N pairs never do. $B$ lies at least $26^{\circ}$ (ax+bas, C–N) or $23^{\circ}$ (bas+bas) from either large computed axis.

*$V_{Ga}^{-}$ with two basal holes.* The state was initialized at the hybrid level (PBE nearly delocalizes its holes) and relaxed under an occupation constraint; an unconstrained force check met the tolerance without moving any ion. Its energy differs from one matched ax+bas pattern by less than the estimated systematic sensitivity. With its tensor in place of the borrowed ones, the earlier populations fail the tolerances; the other basal-pair tensors are its symmetry copies. The needed mean-energy difference may have either sign (the measured tensor's overall sign is open). The single static energy difference does not determine the mean pattern-energy difference or the free-energy difference. The absolute scale of the tensor has not been validated.

*Averaging.* The mirror pair gives $\rho = -0.58$ to $-0.57$, but $+0.39$ with pair tensors. At 295 K, B1+B2 accounts for 0.51–0.59 of the population. Averaging requires an exchange rate above $2\pi$ times the tensor spread, $2.7\text{–}4.1 \times 10^{10}$ s$^{-1}$. An exchange rate of about $0.6\text{–}2.3 \times 10^{12}$ s$^{-1}$ is needed to keep the added width below the measured 60–100 MHz. For an assumed attempt frequency of $10^{13}$ $s^{-1}$, barriers must lie below about 0.04–0.07 eV (computed-tensor spread) or 0.10–0.13 eV (pair spread).

## R17. Further density-functional checks

*$Ga_N^0$.* The spin–spin tensor has $|D| = 2.26$ GHz (2.7 times the measured value), $E/D = 0.04$ and main axis $23^{\circ}$ from $c$.

*Hop barriers.* For $(V_{Ga}\text{-}H)^0$, a five-image nudged elastic band gives 107.5–110.3 meV above A+B1 for the hop into B1+B2 and 118.0–118.5 meV for the mirror hop (highest image to cubic-fit maximum). In a stochastic-Liouville model at 295 K, placing at least 90 % of the intensity in lines as narrow as those measured requires barriers of at most about 0.05–0.08 eV for the computed tensors and an assumed $\nu = 10^{13}$ $s^{-1}$. These barrier thresholds specify the model conditions needed to reproduce the measured widths; the actual hopping rates remain undetermined.

*Larger cell and PAW sets.* In the 192-site cell, where the mirror through $c$ and B0 is a lattice symmetry, A+B1 and B1+B2 of this complex have $|D| = 4.77$ and 5.21 GHz. B1+B2 lies 16.1 meV below the degenerate mirror pair. In the 92-site cell, it lies 17.7 meV below the mean energy of the two computed mirror partners; the 18.3 meV value in Sec. S6 refers to A+B1 alone. At the fixed A+B1 geometry and 760 eV, hard N changes $|D|$ by $-4.7\,\%$, the GW Ga set by $+0.1\,\%$ and the fully hard set by $-4.8\,\%$, with no all-electron reference.

*Fe complexes.* $Fe_{Ga}$ was paired with O or an N vacancy on B0, H antibonding to B0, or a Ga vacancy across B0, each from one PBE+$U$ start. For the first three, the sector order agrees with Ref. [B8]. Only one solution was obtained in each sector, so the lowest energy within a sector remains unestablished. No systematic orbital search was made, spin purity was not demonstrated, and $\alpha$ was not varied.

## R18. Transition-metal centers

*Symmetry.* The Cr, Co and Fe sites deviate from $C_{3v}$ by 0.0024–0.0033 Å rms, with the axis within $0.1^{\circ}$ of $c$. Thus, $E = 0$ apart from strain.

*Route A.* The preset test was to reproduce the $Cr^{4+}$ $|D| = 7.335 \pm 0.002$ GHz [B9] within a factor of 2, axis expected along $c$, $E \approx 0$. The total tensor gives $D = +2.20$ GHz, $11^{\circ}$ from $c$, $E/D = 0.22$. At the $C_{3v}$ sites, 0.3–1.3 GHz of spurious $|E|$ set a noise floor containing the $|E| = 0.255$ GHz of group II. The deficit and the artifact are unexplained; $U$ was not varied. If the 1.193 eV center is $Ti^{2+}$ [B10] or $V^{3+}$ [B11], the test is void. Its single zero-field line and zero-phonon line, far from group II [B1,B9], exclude $Cr_{Ga}^{+}$.

*Route B.* The small cluster has the ion and its four N in the quantum region (def2-TZVP [B23]), 12 capped Ga and 2936–2941 point charges. The large cluster (Cr, Co) adds 12 Ga and 25 N (def2-SVP). The spin–spin coupling (SSC) calculation fails in parallel runs of ORCA 6.1.1, so SSC is included only for the small Cr cluster.

*Cr and Fe.* Cr NEVPT2 gives $D = +7.54$ GHz (+9.51 GHz with SSC) in the small cluster and $+8.49$ GHz in the large cluster; all six blocks give 0.89–1.72 times the measured value. Adding small-cluster SSC to the large cluster gives an estimated 1.43 times, outside $\pm 30\,\%$. The spin–orbit terms, mainly from ${}^3T_2$, cancel to about 7 % of the largest, so the agreement is partly fortuitous. Fe CASSCF gives $D = -0.216$ against $-2.25$ GHz [B12] and misses the measured positive $g$ shift. Intruder states invalidate the NEVPT2 result. Route B is calibrated, to a factor of 2 on one anchor, only for orbitally non-degenerate ions with $D$ from second-order spin–orbit coupling via ligand-field $T_2$ states.

*Co.* All $C_{3v}$ parts of $D$ are positive: $+35.51$ and $+34.33$ GHz (CASSCF, NEVPT2) in the small cluster, $+5.95$ and $+9.24$ GHz in the large. The smallest gap is $1.3$ times the Weyl bound on $\Delta_0$, but falls to $0.79$ times that bound after division by the largest Cr ratio; the comparison fails for $^{59}$Co hyperfine couplings above $0.23$ GHz. $Ni^{3+}$ in GaN ($g'_{\parallel}=2.10$, $g'_{\perp}\approx 4.20$ [B13]) leads us to infer that $2D$ lies far above the X-band quantum, although the margin is unquantified.

*Ni.* $C_3$ breaking ($0.014$ Å rms) splits $^3T_1$ by $46$ and $83$ cm$^{-1}$ (CASSCF) or $20$ and $40$ cm$^{-1}$ (NEVPT2); the next spin–orbit levels lie $8.9$–$11.5$ THz higher.

## R19. Spin–orbit methods

*Route A.* Nine calculations [B14,B15] orient the spin along $x$, $y$, $z$ and six diagonals. Fitting $E(n)=c+n\cdot M\cdot n$ gives $D_{SOC}=M/[S(S-1/2)]$ (spin-coherent states [B16]); LDMATRIX adds $D_{SS}$. Potential biases that were not tested include excitation energies from one $U$; Kohn–Sham weighting of spin-flip excitations [B17–B19]; frozen density; orbital degeneracy (Ni); fourth-rank terms for $S\geq 2$; a cell lacking $C_3$.

*Route B.* We use quasi-degenerate perturbation theory [B24] with the mean-field spin–orbit operator (Sec. S7). Capped Ga carry large-core pseudopotentials [B20]; no bond is cut. Outer point charges are fitted to neutrality, zero dipole and the Ewald potential [B21]. AVAS [B22] selects the Co, Ni and Fe $3d$ orbitals. Biases include formal charges, a finite quantum region and the hybrid geometry.

## R20. Instrument-aware robustness

*Archive.* Besides the maps and defect-6 Rabi data, the archive holds pulsed ODMR, microwave-on/off time-resolved photoluminescence, cw traces near 1 kG and angle scans of defects 1 and 2; species-labeled angle spectra of defect 2; photon correlations of defects 1–4 and 6; and photoluminescence spectra, maps and magneto-photoluminescence. The defect-2 contrast of $c$ is $5.1\%$ in pulsed ODMR (others $\leq 1.6\%$) and $33\%$ near 1 kG (others $4.5$–$7.0\%$; single-point maxima).

*Window models.* B1f (common lower-window field offset) gives $q-(u+l)=-2.2\pm 7.7$ MHz; B3f gives $+0.1\pm 22.8$ MHz, set by its prior box. When $q$ is excluded, the $S=1$ fit requires a lower-window $l$ shift of $-5.4\pm 5.2$ MHz, about $2\sigma$ from the overlap step; the ladder needs $-26.6\pm 4.0$ (B2f) or $+18.7\pm 2.7$ MHz (B2uf). These offsets disfavor the $S=1$ description of $l$.

*Model comparisons.* The $S=3/2[u,l,q]$ model is rejected under every window model ($\Delta BIC\geq 5281$) and every resample ($\geq 193$). Map fits rank the $S=1+c$ model first in $9$ of $10$ field-range and nuisance variants but, on the $q$ range, do not separate it from the ladder with a free quadratic line at $q$ (Cox $p=0.20$). Parsimony favors the $S=1+c$ model ($\Delta AIC=+7.1$, but $-1.3$ without replicas).

*Simulated maps.* Closure near zero, $q-2c$ far from zero and $\Delta\chi^2\geq 7942$ hold together in $0.94$ of maps simulated from the $S=1+c$ model and $0.93$ from the ladder with an independent line at $q$. Alternatives that predict their own $q$ reach this in none of $200$ realizations each ($100$ for one). This primarily checks the analysis on simulated data. The ladder reaches the closure trend in $2.5\%$; the Gaussian $\log_{10}$ likelihood ratio, $S=1+c$ over ladder, is $+0.64$ (closure), $-1.38$ (trend) and $+0.14$ (both).

*Coincidence.* Of the search-box tracks, $3.4\times 10^{-5}$ lie within $10$ MHz rms of $q$. Under B2f the $S=1$ prediction misses $q$ by $-16.0\pm 11.8$ MHz on average.

## R21. Optical-cycle models and low-field positions

The isolated-spin cw drive is treated exactly with a laboratory-frame Floquet calculation. A saturating incoherent microwave rate, a truncated rotating-wave window or a microwave-off reference in another frame can each spuriously reproduce the $q$ amplitude pattern. First-photon pile-up of the transients is modeled exactly; count rates of other sessions set its nuisance range, $0.12$–$0.22$ recorded photons per laser cycle. The data files state neither the laser power, background and count rate of the transient session nor the microwave field at the defect.

For the isolated $S=1$ model, the whole-map statistic lies $1.9$–$2.9$ replicate standard deviations above the replicate mean. The comparisons use $20$ closed-form replicates, $90\,\%$ of which reached the time limit, and $8$ replicates with the exact line shape. These exploratory comparisons do not provide a calibrated statistical significance for the observed deviation. Of the $128$ candidate end points of that search, $127$ give $q/u$ falling from 0 to $+6$ dB; the exception is a rejected fit with $3$ times the best map statistic.

On the $15$ low-field points the ladder of the first search has $\chi^2=84.3$ and $S=1+c$ $20.3$; under the steeper field drift, $11.1$ and $38.4$.

## R22. Cross-checks

Line-extraction and spin-Hamiltonian results were cross-checked over matching field ranges, and electronic-structure quantities were recomputed from the original output files. Agreement in these checks supports numerical consistency but does not validate assumptions shared by the analyses.

## R23. Data and code availability

The analysis materials also include the fitting code for the $S=3/2[u,l,q]$ model, the transition-metal fits and window test, and the Blender scenes used for the structure figures. Omitted are wave functions, charge densities, other large VASP outputs, failed, duplicate or misconfigured embedded-cluster runs, and licensed PAW files.

## References for Part B